\pdfoutput=1
\documentclass[a4paper,american,aps,citeautoscript,floatfix,pdftex,pra,preprintnumbers,showpacs,superscriptaddress,twoside,%
longbibliography,%
reprint,twocolumn,final% add final to see real layout (e.g., no todos anymore)
]{revtex4-1}
\usepackage{amsfonts,amsmath,amssymb}
\usepackage{graphicx}
\usepackage[T1]{fontenc}
\usepackage[utf8]{inputenc}
\usepackage{microtype}
\usepackage[textsize=footnotesize,obeyFinal]{todonotes}
\usepackage{hyperref,hypernat}
\usepackage[todos]{CMImacros}

\newlength{\figwidth}
\newcommand{\cfeldesy}{\affiliation{Center for Free-Electron Laser Science, Deutsches
      Elektronen-Synchrotron DESY, Notkestrasse 85, 22607 Hamburg, Germany}}%
\newcommand{\uhhcui}{\affiliation{The Hamburg Center for Ultrafast Imaging, Universität Hamburg,
      Luruper Chaussee 149, 22761 Hamburg, Germany}}%
\newcommand{\uhhphys}{\affiliation{Department of Physics, Universität Hamburg, Luruper Chaussee 149,
      22761 Hamburg, Germany}}%
\newcommand{\jkemail}{\email{jochen.kuepper@cfel.de}}%
\newcommand{\cmiweb}{\homepage{https://www.controlled-molecule-imaging.org}}%

\begin{document}
\title{Optimized cell geometry for buffer-gas-cooled molecular-beam sources}%
\author{Vijay Singh}\cfeldesy\uhhcui
\author{Amit K.\ Samanta}\cfeldesy
\author{Nils Roth}\cfeldesy\uhhphys
\author{Daniel Gusa}\cfeldesy
\author{Tim~Ossenbrüggen}\cfeldesy
\author{Igor Rubinsky}\cfeldesy\uhhcui
\author{Daniel A.\ Horke}\cfeldesy\uhhcui%
\author{Jochen Küpper}\jkemail\cmiweb\cfeldesy\uhhcui\uhhphys%

\date{\today}%
\begin{abstract}\noindent%
   We have designed, constructed, and commissioned a cryogenic helium buffer-gas source for
   producing a cryogenically-cooled molecular beam and evaluated the effect of different cell
   geometries on the intensity of the produced molecular beam, using ammonia as a test molecule.
   Planar and conical entrance and exit geometries are tested. We observe a three fold enhancement
   in the NH$_3$ signal for a cell with planar-entrance and conical-exit geometry, compared to that
   for a typically used `box'-like geometry with planar entrance and exit. These observations are
   rationalized by flow-field simulations for the different buffer-gas cell geometries. The full
   thermalization of molecules with the helium buffer-gas is confirmed through rotationally-resolved
   REMPI spectra yielding a rotational temperature of 5~K.
\end{abstract}
\maketitle%

\section{Introduction}
\label{sec:introdution}
The production of slow and cold molecular beams with high flux is a prerequisite for many modern
experiments, such as the study of cold reactive collisions~\cite{Henson:Science338:234,
Singh:PRL108:203201, Chang:Science342:98} or Zeeman and Stark
deceleration~\cite{Vanhaecke:PRA75:031402, Bethlem:PRL83:1558}. Furthermore, they are the starting
point for further cooling of molecules to ultracold temperatures via evaporative or sympathetic
cooling with ultracold atoms~\cite{Stuhl:Nature492:396, Lim:PRA92:053419} or by direct laser cooling
of molecules~\cite{Shuman:Nature467:820, Zhelyazkova:PRA89:053416, Kozyryev:PRL118:173201}. These
novel sources of ultracold molecular systems enable ultrahigh precision measurements, such as the
investigation of spatial and temporal variations of fundamental
constants~\cite{Santamaria:JMolSpec300:116, Truppe:NatComm4:2600}, and experimental searches of the
electric dipole moment of the electron~\cite{Hudson:Nature473:493, ACME:Science343:269}.

The advent of ultrashort x-ray free-electron lasers has now sparked an immense interest in producing
cold beams of large molecules and particles for x-ray diffractive imaging with atomic
resolution~\cite{Seibert:Nature470:78, Ekeberg:PRL114:098102, Barty:ARPC64:415,
Kuepper:PRL112:083002, Filsinger:PCCP13:2076}. These experiments require pure samples delivered into
a micrometer-sized interaction region, which can be achieved using electrostatic control techniques
applied to cold molecular beams~\cite{Filsinger:PRL100:133003, Filsinger:PCCP13:2076,
Chang:IRPC34:557}. Moreover, these techniques also allow the dispersion of rotational states, such
that the coldest molecules from a molecular beam can be selected, yielding, \eg, higher degrees of
molecular one- and three-dimensional alignment and orientation~\cite{Holmegaard:PRL102:023001,
Filsinger:JCP131:064309, Nevo:PCCP11:9912, Trippel:MP111:1738, Trippel:PRA89:051401R,
Trippel:PRL114:103003}.

All the aforementioned experiments, from laboratory-based high-precision spectroscopy to
facility-based novel imaging methods, require stable sources producing a high flux of cold molecules
in the gas-phase. One approach to achieve this is the use of buffer-gas cooling
techniques~\cite{Hutzler:CR112:4803, Truppe:JModOpt65:246}, in which a molecular
sample is cooled through collisions with a cryogenic coolant, typically helium. In this paper we
report on our newly developed cryogenic buffer-gas-cell source, including the geometric optimization
of the cell shape to avoid molecular losses within the cell and produce a higher-flux molecular
beam.

To the best of our knowledge, this represents the first study optimizing the cryogenic cell shape
for the production of higher-flux molecular beams. Our experimental measurement together with
detailed flow-field simulations suggest that a cryogenic cell with a planar entrance and exit, \ie,
the typical ``box''-like geometry, produces vortices in the flow fields, resulting in trapping of
molecules, and eventually loss of molecular density by diffusion to the cold cell walls. We find
that this can be avoided, and a three times higher molecular flux achieved, through the use of a
conical-exit aperture. By using ammonia as a test system we confirm that the altered geometry does
not affect the thermalization of molecules, which emerge from the cell with a rotational temperature
of $\ordsim5$~K.

\section{Experimental Methods}
\label{sec:methods}
\begin{figure}
   \centering%
   \includegraphics[width=\linewidth]{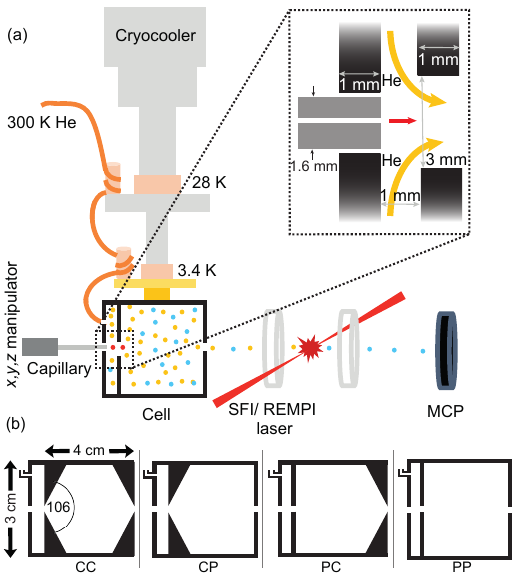}
   \caption{(a) Schematic of the cryogenic buffer-gas cell setup. A two-stage cryocooler holds a
      copper cell, cooled to 3.4~K base temperature. The cell has a small volume at the front where
      cold helium (orange dots) is introduced, which then flows radially into the main
      thermalization cell. Warm molecules (red dots) are injected from a heated stainless steel
      capillary, carried by the radial helium flow and eventually thermalize to the buffer-gas
      temperature (blue dots) via collisions. The produced molecular beam is detected in a
      time-of-flight mass spectrometer, see text for details. (b) The buffer-gas cell has detachable
      entrance and exit end caps, which can either be planar, with an \degree{180} opening angle, or
      conical, with an \degree{106} opening angle, resulting in four distinct geometries.}
   \label{fig:cell}
\end{figure}
A schematic diagram of our molecular beam apparatus is shown in~\subautoref{fig:cell}{a}. It
primarily consists of a buffer-gas cell source cooled by a pulsed-tube refrigerator (Sumitomo
RP082E2) and a time-of-flight mass spectrometer (TOF-MS) with a multi-channel plate (MCP) detector.
These are mounted in vacuum chambers evacuated using turbomolecular pumps to typical operating
pressures of 2$\times10^{-6}$~mbar (source region, Pfeiffer Vacuum HiPace 2300) and
8$\times10^{-7}$~mbar (detection region, Pfeiffer Vacuum HiPace 300).

The cryocooler contains two cooling stages, cooled to base temperatures of 28~K and 3.4~K,
respectively. Each cooling stage is encased by a radiation shield, not shown
in~\subautoref{fig:cell}{a}, to avoid unwanted heat load from black-body radiation. The outer
radiation shield is made of aluminum and thermally connected to the first stage of the cryocooler.
The inner radiation shield is made of oxygen-free copper and thermally connected to the second
stage. The inside of this shield is furthermore covered with coconut charcoal acting as a cryogenic
sorption pump when cooled below 10~K~\cite{Tobin:JVSTA5:101}. Both radiation shields contain 1~cm
diameter apertures at the entrance and exit side to allow introduction and extraction of sample from
the buffer-gas cell, \emph{vide infra}. The buffer-gas cell itself is attached to the second stage
of the cryocooler and cooled down to 3.4~K base temperature. The temperatures at the first and
second cooling stage and at the buffer-gas cell are continuously monitored using calibrated silicon
diode temperature sensors (DT-670, Lake Shore Cryotronics).

The main body of the buffer-gas cell is a hollow copper cylinder of diameter 3~cm and length 2~cm.
Attach to this are exchangeable copper end caps at the entrance and exit side, each 1~cm deep and
with entrance and exit aperture diameters of 3~mm and 2~mm, respectively. In this study we test two
different end-cap geometries for the entrance and exit side of the cell; a common planar end cap
with a \degree{180} opening angle, and a conical end cap with an opening angle of \degree{106}, as
indicated in \subautoref{fig:cell}{b}. This leads to four different geometries of the buffer-gas
cell, depending on the combination of entrance and exit end cap: conical-conical (CC),
conical-planar (CP), planar-conical (PC), and planar-planar (PP).

Helium buffer-gas is introduced into the cell from the same side as the molecular sample, left hand
side in \subautoref{fig:cell}{a}, using a 1.64~mm inner diameter (ID) copper tube. This is thermally
anchored using brazed copper bobbins to both cooling stages to cool the helium to 3.4~K before it
enters the cell. The helium is introduced into a small volume before the actual buffer-gas cell
which it rapidly fills and then provides a radial helium flow into the cell through the 3~mm
entrance end cap aperture, see inset in \subautoref{fig:cell}{a}. The exact flow of helium into the
cell is controlled using a digital flow meter (Vögtlin Instruments GSC-A9TA-BB21) calibrated for
helium and situated outside vacuum. The warm molecular sample, NH$_3$ with purity $>99.99\%$ and
kept at 6~mbar pressure, is introduced into the buffer-gas cell using a 10~cm long stainless steel
capillary with an ID of 254~\um. This capillary is connected to the sample delivery manifold by 6~mm
stainless steel tubing and the complete assembly is attached to the vacuum chamber using a
three-dimensional ($X,Y,Z$) position manipulator. This allows precise manipulation of the capillary
in and out of the buffer-gas cell through the radiation shields and cell-entrance apertures. During
experiments the capillary tip is located around 1~mm inside the cell-entrance aperture, as indicated
in \subautoref{fig:cell}{a}. Molecules are then carried into the cell by the flow of cold helium. To
prevent freezing of sample inside the capillary, it is heated to~\celsius{80} using resistive
heating wire and the temperature is continuously monitored with a thermocouple. For NH$_3$ pressures
of $\le6$~mbar this prevents capillary clogging and the experiment runs continuously for an entire
day, while for NH$_3$ pressure of $\ge10$~mbar clogging occurs after around 3 hours for the used
capillary position. Within the cell warm molecules collisionally thermalize with the cold helium,
before being extracted through an exit aperture of 2~mm diameter into high vacuum. The formed
molecular beam passes through the apertures in both radiation shields and enters into the ion optics
of the TOFMS, located 51~mm downstream of the buffer-gas cell exit.

We can estimate the approximate density $n_\text{He}$ of helium buffer-gas in the cell by
considering a steady-state with a constant and controllable helium flow into the cell, $f$, and a
corresponding flow through the exit aperture with area $A_\text{aperture}$,
\begin{equation}
   n_\text{He} = \frac{4f}{A_\text{aperture}\,\bar{v}}.
   \label{eq:density}
\end{equation}
Here $\bar{v}$ is the mean thermal velocity of helium buffer-gas near the exit aperture, which can
be evaluated as
\begin{equation}
  \bar{v} =\sqrt{\frac{8k_\text{B}T}{\pi{}m}},
\end{equation}
where $k_\text{B}$ is the Boltzmann constant, $T_\text{He}$ the buffer-gas temperature, and
$m_\text{He}$ the mass of the buffer-gas. For a typical flow of $10$~\sccm~\footnote{We use
   ``milliliter normal per minute'' as volume- equivalent of mass flow, with standard temperature
   and pressure conditions of \celsius{0} and 1.013~bar; this is identical to the often used literal
   specification ``sccm''.} into the buffer-gas cell and an exit aperture of 2~mm, this corresponds
to a density of 4$\times10^{16}$~cm$^{-3}$ inside the cell. We furthermore simulate the helium flow
field inside the buffer-gas cell using a finite-element solver, COMSOL Multiphysics with laminar
flow interface~ \cite{Comsol:Multiphysics:5.3}, treating helium as an ideal gas at 4~K temperature.

Between the ion optics of the TOF-MS spectrometer, molecules are ionized by a pulsed laser system.
To monitor the total molecular flux we utilize strong-field ionization (SFI) by an amplified
Ti:Sapphire laser (Spectra Physics Spitfire Ace) yielding 40~fs pulses at 1~kHz repetition rate,
focused to typical field strengths of $10^{13}\text{~W/cm}^2$ in a 100~\um spot (FWHM) by a
$f=750$~mm plano-convex lens. For rotational-state selective detection we use resonance-enhanced
multiphoton ionization (REMPI) of ammonia via the $\tilde{C}\leftarrow\tilde{X}$ transition at
63846--63919~\invcm~\cite{Nolde:PCCP7:1527}. Narrowband laser pulses are provided by a tunable dye
laser (FineAdjustment) using DCM dye, pumped by the frequency doubled output of a Nd:YAG laser
(InnoLas). Typical pulse energies are 2~mJ in a 100~\um spot (FWHM) at 20~Hz repetition rate.
Produced ions are subsequently detected using a time-of-flight mass spectrometer with a typical mass
resolution $m/\Delta{m}\approx1000$.

\section{Results and Discussion}
\label{sec:discussion}
\begin{figure}
   \centering%
   \includegraphics[width=\linewidth]{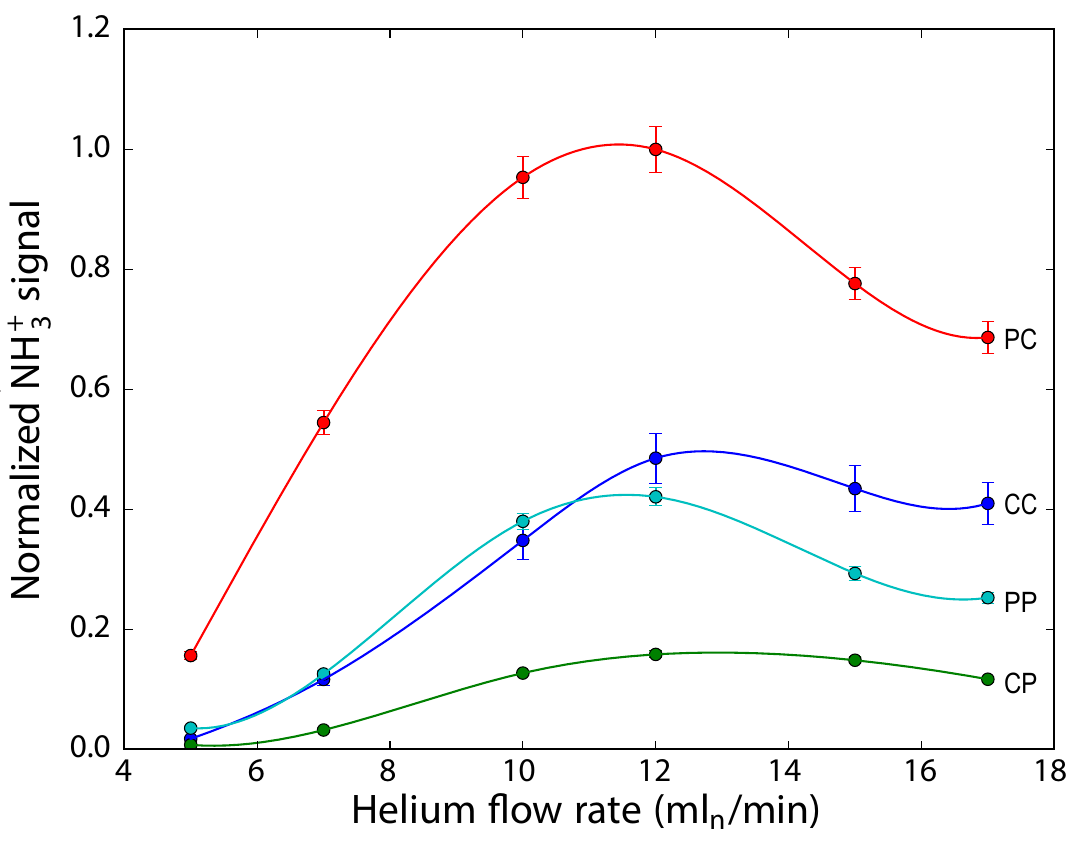}
   \caption{Measured ammonia-ion signal (solid circles), from SFI of ammonia, as a function of
      helium flow rate for the four geometries of the cryogenic cell; the signal
         decreases from PC over CC and PP, to CP. Lines are third-order polynomial fits to guide
      the eye. All geometries show an initial linear increase, until a maximum is reached around
      12~\sccm helium flow. The cell geometry significantly affects the measured signal intensity,
      with a planar-conical cell providing the most intense molecular beam.}
   \label{fig:signal}
\end{figure}
In \autoref{fig:signal} we show the integrated NH$_3^+$ signal for the four different buffer-gas
cell geometries, following strong-field ionization, as a function of the helium flow rate into the
cell. Regardless of the cell geometry, we observe the same general trend: the ammonia signal first
increases approximately linearly with helium flow rate until it reaches a maximum at around
10--12~\sccm of helium, after which the signal plateaus off or decreases slowly.

The initial near-linear rise of the signal with increasing helium flow can be rationalized with the
more efficient extraction of molecules from the buffer-gas cell. This extraction efficiency is often
expressed in terms of the cell-extraction parameter, $\gamma_\text{e}$, defined as the ratio of the
diffusion time $\tau_\text{diff}$ and the typical cell pump-out time
$\tau_\text{pump}$~\cite{Hutzler:CR112:4803}:
\begin{equation}
   \gamma_\text{e} = \frac{\tau_\text{diff}}{\tau_\text{pump}} = \frac{4}{9\pi}
   \frac{n_\text{He}\sigma A_\text{aperture}}{l_\text{cell}} \approx \frac{\sigma f}{l_\text{cell}\bar{v}}.
  \label{eq:extraction}
\end{equation}
Here $\sigma$ is the elastic collision cross-section for molecule-helium collisions, $l_\text{cell}$
the cell length and $f$ is the controllable helium flow into the cell. For $\gamma_\text{e}\leq1$
extraction from the cell is diffusion limited and increased helium flow leads to a corresponding
increase in extraction efficiency, and hence higher signals. For the case where
$\gamma_\text{e}\geq1$, however, most molecules are extracted from the cell before diffusing to the
cell walls, in a regime termed hydrodynamic enhancement. Once this regime is reached an increase in
helium flow rate does not have a significant influence on the extraction efficiency anymore. For our
cell, an extraction parameter of $\gamma_\text{e}\approx1$ is expected at around 13~\sccm flow rate,
assuming a collision cross-section of $\sigma$ = 10$^{-14}$~cm$^2$~\cite{Hutzler:CR112:4803}. This
correlates well with the observed signal emerging from the buffer-gas cell: For all geometries, we
first observe a near-linear increase with helium flow rate, corresponding to the diffusion-limited
case, until the flow rate reaches around 10--12~\sccm. Beyond this we enter the hydrodynamic
entrainment regime, where the signal plateaus off or decreases slowly~\cite{Patterson:JCP126:154307,
   Hutzler:CR112:4803}\footnote{One possible loss mechanism of NH$_3$ signal is the
      formation of dimers or clusters via molecule-molecule collisions near the cell entrance.
      However, our NH$_3$ signal is observed to increase linearly with NH$_3$ inlet pressure and we
      rule out the formation of dimers or clusters and hence NH$_3$ loss by this mechanism.}.

\begin{figure}
   % moved here in order to get it onto page 3 -- so this one is next to cap-dependent signals and
   % the subsequent floats can be placed nicely
   \centering%
   \includegraphics[width=\linewidth]{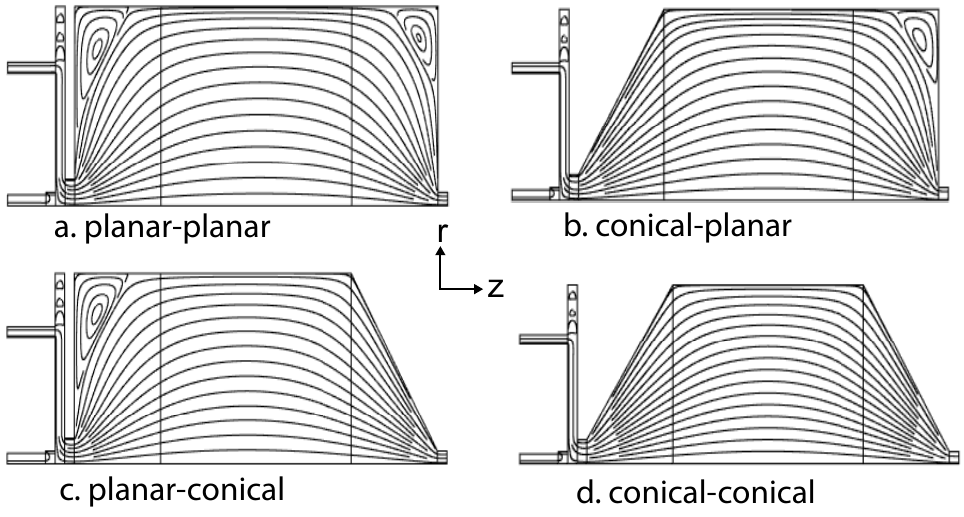}
   \caption{Simulations of the helium flow-field for four different geometries of our cryogenic
   cell. Depicted are streamlines with the helium flowing from left to right. For clarity, only one
   half of the cylindrically-symmetric $r,z$-plane is shown.}
   \label{fig:simulation}
\end{figure}

While the general trend between the observed ammonia signal and the applied helium flow is similar
for all cell geometries tested, we do observe marked differences in the absolute signal levels
obtained for different cell configurations, as shown in~\autoref{fig:signal}. The cell with a planar
entrance and a conical-exit end cap (PC) produces the largest signal, with about a threefold
increase compared to the regular planar-planar (PP) shape. In contrast to this, the cell with a
conical-entrance and planar-exit end cap (CP) produces the smallest signal levels, almost tenfold
smaller than the PC configuration.

In order to qualitatively understand the observed differences for the four cell geometries, we have
simulated the helium flow fields in the cryogenic cell, see~\autoref{sec:methods} for details, at
10~\sccm of helium flow and a pressure of $10^{-6}$~mbar at the cell exit. The resulting flow fields
are shown in \autoref{fig:simulation} as slices through the cylindrically-symmetric cell volume,
\ie, in $r,z$ space, and we further extract the helium pressures inside the cell as well as the flow
of helium through the entrance and exit of the cell, shown in \autoref{tab:simulation}.
\begin{table*}
   \caption{Simulated pressures and helium flows within the cell for the four geometries at 10~\sccm
      helium flow rate.}
   \begin{tabular}{ccccc}
     \hline\hline
     geometry	& pressure (mbar)& density (cm$^{-3}$)	& outlet flow (\sccm) & inlet flow (\sccm)\\
     \hline
     CC&0.033&7.9$\times$10$^{16}$&9.6 &0.4 \\
     CP&0.048&1.2$\times$10$^{17}$&9.2&0.8 \\
     PC&0.032&7.7$\times$10$^{16}$ &9.6&0.4\\
     PP & 0.048&1.2$\times$10$^{17}$ &9.2& 0.8 \\
     \hline
   \end{tabular}
   \label{tab:simulation}
\end{table*}

These flows and cell pressures clearly show that the helium environment within the cell depends on
the geometry of the exit end cap. A planar-exit end cap leads to reduced transmission through the
exit aperture compared to a conical end cap, yielding a correspondingly higher pressure within the
cell and loss of helium density by `back-flow' through the cell-entrance aperture. Furthermore, the
simulations show that planar end caps produce vortices in the flow fields at the cell corners. This
can be circumvented by the use of conical end caps, which result in an overall laminar flow. Hot
ammonia molecules introduced into the cell will be thermalized with the buffer-gas before reaching
the exit end cap and hence follow the streamlines. The presence of vortices in the flow-field can,
therefore, lead to trapping of ammonia and eventually loss to the cell walls. Thus, the laminar flow
produced by a conical end cap at the cell exit is clearly advantageous in avoiding diffusion loss
and, therefore, entraining more NH$_3$ molecules into the molecular beam.

Furthermore, according to our calculations the effect of the end caps on the forward velocity of the
generated beams is small, but when replacing the planar exit end cap by the conical exit end cap it
decreases from 80~m/s to 70~m/s, which is advantageous for slow-beam and trapping experiments.

The simulation results for the extracted helium flow and cell pressures for CC and PC cell
geometries are very similar, indicating that the vortices produced by the planar end
cap at the cell entrance do not have any adverse influence on the helium flow through the cell. This
might lead one to expect very similar ammonia signal levels for these two cell geometries. However,
the measured signal levels differ by around a factor of 3. We attribute this difference to increased
diffusion and loss of ammonia molecules at the cold cell walls for the conical-entrance cap due to
the closer proximity of the copper walls to the inlet capillary. At the cell entrance molecules are
not fully thermalized with the helium buffer-gas and will not follow the helium flow-field
precisely. Therefore, diffusive motion is still considerable and the larger opening angle of the
planar-entrance cap leads to a larger distance to the cold cell walls and hence an increased
molecular flux out of the buffer-gas cell compared to the conical-entrance geometry.

Based on these arguments and our flow-field simulations, we summarize our observation as follows: at
the entrance molecules are primarily lost through diffusion, making planar end caps preferential as
they maximize the distance to the cold cell walls. At the cell exit molecules are thermalized and
follow the helium flow-field such that vortices should be avoided, which can be achieved with
conical end caps. For our cell, these two effects appear to be of roughly the same magnitude, such
that the planar-planar (PP) and conical-conical (CC) geometries show approximately the same signal
intensity.
\begin{figure}
   \centering%
   \includegraphics[width=\linewidth]{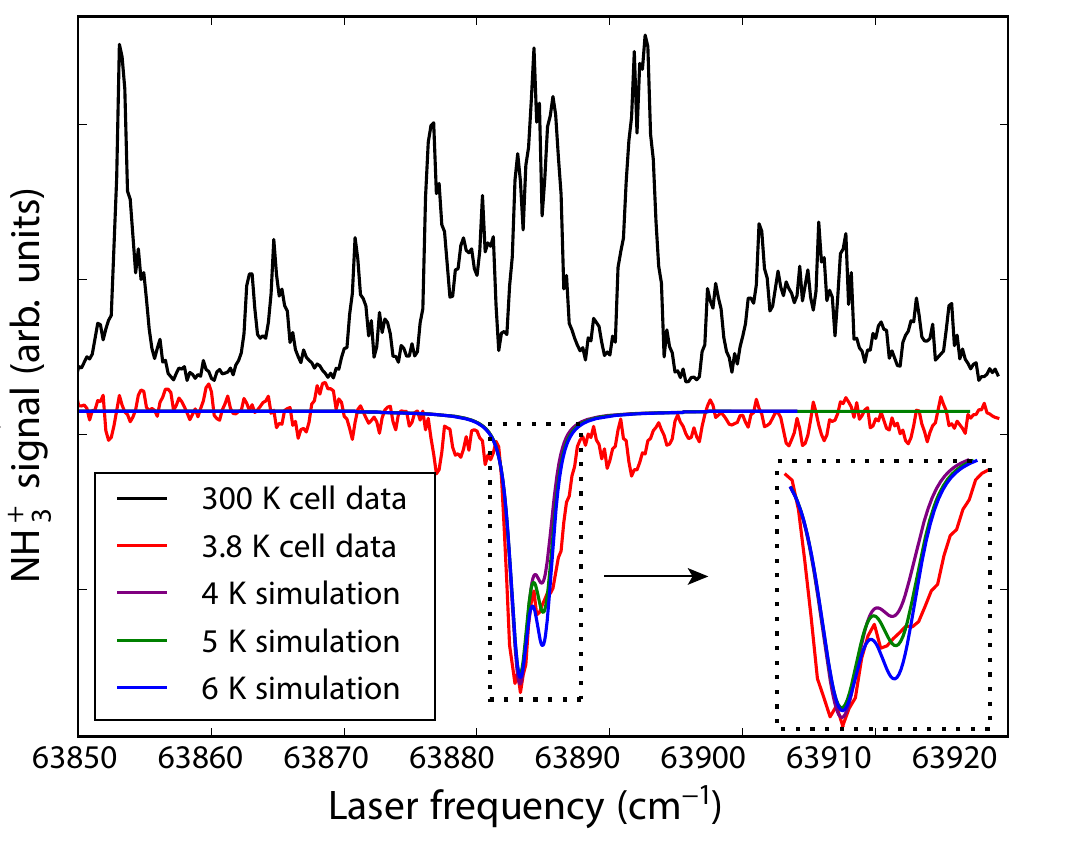}
   \caption{REMPI spectrum of NH$_3$ measured using a planar-conical cell at room temperature
      (black, upper line) and cooled to 3.8~K (red, lower line),
      together with corresponding PGOPHER simulations at 4~K, 5~K, and 6~K rotational temperature.
      In the inset, the different simulations can be distinguished on the right peak,
         which is strongest for 6~K, intermediate for 5~K, and weakest for 4~K. The 5~K simulation
      reproduces the experimental data very well, indicating full thermalization of the molecules
      with the buffer-gas.}
   \label{fig:rempi}
\end{figure}
To confirm that molecules ejected from the cell are fully thermalized with the buffer-gas, we
collected $2+1$~resonance-enhanced multi-photon ionization (REMPI) spectra of ammonia \emph{via} the
$\tilde{C}$ state. Here, an NH$_3$ pressure of 6~mbar and a helium flow rate of
   10~\sccm were used, corresponding to a Reynolds number of $\ordsim56$ and thus producing a
   molecular beam in the partially-hydrodynamic regime, \ie, without any significant further
   acceleration or cooling when exiting the cell~\cite{Hutzler:CR112:4803}. In \autoref{fig:rempi}
the spectrum is shown for both, room temperature and buffer gas cooled NH$_3$. The strong
temperature dependence of the rovibronic transitions allows us to extract precise information about
the rotational temperature of ammonia. We simulated spectra using the PGOPHER
software~\cite{Western:pgopher}, with literature values used for the rotational constants in the
electronic ground ($\tilde{X}$) and excited ($ \tilde{C}$) states~\cite{Nolde:PCCP7:1527}. The
simulated spectrum for a rotational temperature of 5~K is shown as a solid green line in
\autoref{fig:rempi}. It matches the experimental data for buffer-gas cooled NH$_3$ very well. As the
intensities of the two observable transitions ($J=0,1$) are very sensitive to the rotational
temperature, we also simulated spectra at 4~K and 6~K (purple and blue lines in \autoref{fig:rempi},
respectively). These do not fit the experimental spectrum well and confirm our temperature
assignment of 5~K. This indicates complete thermalization of the molecular rotations with the helium
buffer-gas. Since the elastic collision cross-section of the molecules with the helium is expected
to be significantly higher than the that for the rotational relaxation~ \cite{Hutzler:CR112:4803},
the translational temperature of the molecules should also be thermalized completely. Measurements
were conducted for all four cell geometries and similar rotational temperatures extracted in all
cases.

\section{Conclusion}
\label{sec:conclusion}
In conclusion, we designed, constructed, and commissioned a cryogenic apparatus for producing helium
buffer-gas cooled molecular beams. We evaluated the effect of the cell geometry on the produced
molecular beam flux and observed that a cryogenic cell with a planar-entrance and conical-exit end
cap produces around three times more flux compared to the typically used `box'-like geometry with
planar-entrance and exit. This is attributed to the smaller diffusion loss at the entrance due to
the larger opening angle of a planar end cap and an optimum transmission at the exit resulting from
purely laminar helium flow without vortices for the conical end cap. We confirmed the full
thermalization of molecules with the buffer-gas through REMPI measurements that show a rotational
temperature of 5~K for all cell geometries.

We showed that a simple geometric optimization of the cell shape can significantly improve the
molecular beam flux and we are currently expanding this technique to larger molecules and
nanoparticles. Such a stable, cold, and high-flux beam of molecules will be beneficial to a
wide range of molecular physics experiments, including novel single-particle x-ray diffractive
imaging approaches.

\vspace*{1em}

\begin{acknowledgments}\noindent%
   In addition to DESY, this work has been supported by the European Research Council under the
   European Union's Seventh Framework Programme (FP7/2007-2013) through the Consolidator Grant
   COMOTION (ERC-614507-Küpper), by the excellence cluster ``The Hamburg Center for Ultrafast
   Imaging -- Structure, Dynamics and Control of Matter at the Atomic Scale'' of the Deutsche
   Forschungsgemeinschaft (CUI, DFG-EXC1074), and by the Helmholtz Gemeinschaft through the
   ``Impuls- und Vernetzungsfond''.
\end{acknowledgments}
\bibliography{string,cmi}

\begin{thebibliography}{36}%
\makeatletter
\providecommand \@ifxundefined [1]{%
 \@ifx{#1\undefined}
}%
\providecommand \@ifnum [1]{%
 \ifnum #1\expandafter \@firstoftwo
 \else \expandafter \@secondoftwo
 \fi
}%
\providecommand \@ifx [1]{%
 \ifx #1\expandafter \@firstoftwo
 \else \expandafter \@secondoftwo
 \fi
}%
\providecommand \natexlab [1]{#1}%
\providecommand \enquote  [1]{``#1''}%
\providecommand \bibnamefont  [1]{#1}%
\providecommand \bibfnamefont [1]{#1}%
\providecommand \citenamefont [1]{#1}%
\providecommand \href@noop [0]{\@secondoftwo}%
\providecommand \href [0]{\begingroup \@sanitize@url \@href}%
\providecommand \@href[1]{\@@startlink{#1}\@@href}%
\providecommand \@@href[1]{\endgroup#1\@@endlink}%
\providecommand \@sanitize@url [0]{\catcode `\\12\catcode `\$12\catcode
  `\&12\catcode `\#12\catcode `\^12\catcode `\_12\catcode `\%12\relax}%
\providecommand \@@startlink[1]{}%
\providecommand \@@endlink[0]{}%
\providecommand \url  [0]{\begingroup\@sanitize@url \@url }%
\providecommand \@url [1]{\endgroup\@href {#1}{\urlprefix }}%
\providecommand \urlprefix  [0]{URL }%
\providecommand \Eprint [0]{\href }%
\providecommand \doibase [0]{http://dx.doi.org/}%
\providecommand \selectlanguage [0]{\@gobble}%
\providecommand \bibinfo  [0]{\@secondoftwo}%
\providecommand \bibfield  [0]{\@secondoftwo}%
\providecommand \translation [1]{[#1]}%
\providecommand \BibitemOpen [0]{}%
\providecommand \bibitemStop [0]{}%
\providecommand \bibitemNoStop [0]{.\EOS\space}%
\providecommand \EOS [0]{\spacefactor3000\relax}%
\providecommand \BibitemShut  [1]{\csname bibitem#1\endcsname}%
\let\auto@bib@innerbib\@empty
%</preamble>
\bibitem [{\citenamefont {Henson}\ \emph {et~al.}(2012)\citenamefont {Henson},
  \citenamefont {Gersten}, \citenamefont {Shagam}, \citenamefont {Narevicius},\
  and\ \citenamefont {Narevicius}}]{Henson:Science338:234}%
  \BibitemOpen
  \bibfield  {author} {\bibinfo {author} {\bibfnamefont {A~B}\ \bibnamefont
  {Henson}}, \bibinfo {author} {\bibfnamefont {S}~\bibnamefont {Gersten}},
  \bibinfo {author} {\bibfnamefont {Y}~\bibnamefont {Shagam}}, \bibinfo
  {author} {\bibfnamefont {J}~\bibnamefont {Narevicius}}, \ and\ \bibinfo
  {author} {\bibfnamefont {E}~\bibnamefont {Narevicius}},\ }\bibfield  {title}
  {\enquote {\bibinfo {title} {Observation of resonances in penning ionization
  reactions at sub-{K}elvin temperatures in merged beams},}\ }\href {\doibase
  10.1126/science.1229141} {\bibfield  {journal} {\bibinfo  {journal}
  {Science}\ }\textbf {\bibinfo {volume} {338}},\ \bibinfo {pages} {234--238}
  (\bibinfo {year} {2012})}\BibitemShut {NoStop}%
\bibitem [{\citenamefont {Singh}\ \emph {et~al.}(2012)\citenamefont {Singh},
  \citenamefont {Hardman}, \citenamefont {Tariq}, \citenamefont {Lu},
  \citenamefont {Ellis}, \citenamefont {Morrison},\ and\ \citenamefont
  {Weinstein}}]{Singh:PRL108:203201}%
  \BibitemOpen
  \bibfield  {author} {\bibinfo {author} {\bibfnamefont {Vijay}\ \bibnamefont
  {Singh}}, \bibinfo {author} {\bibfnamefont {Kyle~S.}\ \bibnamefont
  {Hardman}}, \bibinfo {author} {\bibfnamefont {Naima}\ \bibnamefont {Tariq}},
  \bibinfo {author} {\bibfnamefont {Mei-Ju}\ \bibnamefont {Lu}}, \bibinfo
  {author} {\bibfnamefont {Aja}\ \bibnamefont {Ellis}}, \bibinfo {author}
  {\bibfnamefont {Muir~J.}\ \bibnamefont {Morrison}}, \ and\ \bibinfo {author}
  {\bibfnamefont {Jonathan~D.}\ \bibnamefont {Weinstein}},\ }\bibfield  {title}
  {\enquote {\bibinfo {title} {Chemical reactions of atomic lithium and
  molecular calcium monohydride at 1 {K}},}\ }\href {\doibase
  10.1103/PhysRevLett.108.203201} {\bibfield  {journal} {\bibinfo  {journal}
  {Phys. Rev. Lett.}\ }\textbf {\bibinfo {volume} {108}},\ \bibinfo {pages}
  {203201} (\bibinfo {year} {2012})}\BibitemShut {NoStop}%
\bibitem [{\citenamefont {Chang}\ \emph {et~al.}(2013)\citenamefont {Chang},
  \citenamefont {D{\l}ugo\l\k{e}cki}, \citenamefont {K{\"u}pper}, \citenamefont
  {R{\"o}sch}, \citenamefont {Wild},\ and\ \citenamefont
  {Willitsch}}]{Chang:Science342:98}%
  \BibitemOpen
  \bibfield  {author} {\bibinfo {author} {\bibfnamefont {Y.-P.}\ \bibnamefont
  {Chang}}, \bibinfo {author} {\bibfnamefont {K.}~\bibnamefont
  {D{\l}ugo\l\k{e}cki}}, \bibinfo {author} {\bibfnamefont {J.}~\bibnamefont
  {K{\"u}pper}}, \bibinfo {author} {\bibfnamefont {D.}~\bibnamefont
  {R{\"o}sch}}, \bibinfo {author} {\bibfnamefont {D.}~\bibnamefont {Wild}}, \
  and\ \bibinfo {author} {\bibfnamefont {S.}~\bibnamefont {Willitsch}},\
  }\bibfield  {title} {\enquote {\bibinfo {title} {Specific chemical
  reactivities of spatially separated 3-aminophenol conformers with cold
  {Ca$^+$} ions},}\ }\href {\doibase 10.1126/science.1242271} {\bibfield
  {journal} {\bibinfo  {journal} {Science}\ }\textbf {\bibinfo {volume}
  {342}},\ \bibinfo {pages} {98--101} (\bibinfo {year} {2013})},\ \Eprint
  {http://arxiv.org/abs/1308.6538} {arXiv:1308.6538 [physics]} \BibitemShut
  {NoStop}%
\bibitem [{\citenamefont {Vanhaecke}\ \emph {et~al.}(2007)\citenamefont
  {Vanhaecke}, \citenamefont {Meier}, \citenamefont {Andrist}, \citenamefont
  {Meier},\ and\ \citenamefont {Merkt}}]{Vanhaecke:PRA75:031402}%
  \BibitemOpen
  \bibfield  {author} {\bibinfo {author} {\bibfnamefont {N.}~\bibnamefont
  {Vanhaecke}}, \bibinfo {author} {\bibfnamefont {U.}~\bibnamefont {Meier}},
  \bibinfo {author} {\bibfnamefont {M.}~\bibnamefont {Andrist}}, \bibinfo
  {author} {\bibfnamefont {B.~H.}\ \bibnamefont {Meier}}, \ and\ \bibinfo
  {author} {\bibfnamefont {F.}~\bibnamefont {Merkt}},\ }\bibfield  {title}
  {\enquote {\bibinfo {title} {Multistage {Z}eeman deceleration of hydrogen
  atoms},}\ }\href {\doibase 10.1103/PhysRevA.75.031402} {\bibfield  {journal}
  {\bibinfo  {journal} {Phys.\ Rev.\ A}\ }\textbf {\bibinfo {volume} {75}},\
  \bibinfo {pages} {031402(R)} (\bibinfo {year} {2007})}\BibitemShut {NoStop}%
\bibitem [{\citenamefont {Bethlem}\ \emph {et~al.}(1999)\citenamefont
  {Bethlem}, \citenamefont {Berden},\ and\ \citenamefont
  {Meijer}}]{Bethlem:PRL83:1558}%
  \BibitemOpen
  \bibfield  {author} {\bibinfo {author} {\bibfnamefont {H.~L.}\ \bibnamefont
  {Bethlem}}, \bibinfo {author} {\bibfnamefont {G.}~\bibnamefont {Berden}}, \
  and\ \bibinfo {author} {\bibfnamefont {G.}~\bibnamefont {Meijer}},\
  }\bibfield  {title} {\enquote {\bibinfo {title} {Decelerating neutral dipolar
  molecules},}\ }\href {\doibase 10.1103/PhysRevLett.83.1558} {\bibfield
  {journal} {\bibinfo  {journal} {Phys.\ Rev.\ Lett.}\ }\textbf {\bibinfo
  {volume} {83}},\ \bibinfo {pages} {1558--1561} (\bibinfo {year}
  {1999})}\BibitemShut {NoStop}%
\bibitem [{\citenamefont {Stuhl}\ \emph {et~al.}(2012)\citenamefont {Stuhl},
  \citenamefont {Hummon}, \citenamefont {Yeo}, \citenamefont
  {Qu{\'{e}}m{\'{e}}ner}, \citenamefont {Bohn},\ and\ \citenamefont
  {Ye}}]{Stuhl:Nature492:396}%
  \BibitemOpen
  \bibfield  {author} {\bibinfo {author} {\bibfnamefont {Benjamin~K.}\
  \bibnamefont {Stuhl}}, \bibinfo {author} {\bibfnamefont {Matthew~T.}\
  \bibnamefont {Hummon}}, \bibinfo {author} {\bibfnamefont {Mark}\ \bibnamefont
  {Yeo}}, \bibinfo {author} {\bibfnamefont {Goulven}\ \bibnamefont
  {Qu{\'{e}}m{\'{e}}ner}}, \bibinfo {author} {\bibfnamefont {John~L.}\
  \bibnamefont {Bohn}}, \ and\ \bibinfo {author} {\bibfnamefont {Jun}\
  \bibnamefont {Ye}},\ }\bibfield  {title} {\enquote {\bibinfo {title}
  {Evaporative cooling of the dipolar hydroxyl radical},}\ }\href {\doibase
  10.1038/nature11718} {\bibfield  {journal} {\bibinfo  {journal} {Nature}\
  }\textbf {\bibinfo {volume} {492}},\ \bibinfo {pages} {396--400} (\bibinfo
  {year} {2012})}\BibitemShut {NoStop}%
\bibitem [{\citenamefont {Lim}\ \emph {et~al.}(2015)\citenamefont {Lim},
  \citenamefont {Frye}, \citenamefont {Hutson},\ and\ \citenamefont
  {Tarbutt}}]{Lim:PRA92:053419}%
  \BibitemOpen
  \bibfield  {author} {\bibinfo {author} {\bibfnamefont {Jongseok}\
  \bibnamefont {Lim}}, \bibinfo {author} {\bibfnamefont {Matthew~D.}\
  \bibnamefont {Frye}}, \bibinfo {author} {\bibfnamefont {Jeremy~M.}\
  \bibnamefont {Hutson}}, \ and\ \bibinfo {author} {\bibfnamefont {M.~R.}\
  \bibnamefont {Tarbutt}},\ }\bibfield  {title} {\enquote {\bibinfo {title}
  {Modeling sympathetic cooling of molecules by ultracold atoms},}\ }\href
  {\doibase 10.1103/PhysRevA.92.053419} {\bibfield  {journal} {\bibinfo
  {journal} {Phys. Rev. A}\ }\textbf {\bibinfo {volume} {92}},\ \bibinfo
  {pages} {053419} (\bibinfo {year} {2015})}\BibitemShut {NoStop}%
\bibitem [{\citenamefont {Shuman}\ \emph {et~al.}(2010)\citenamefont {Shuman},
  \citenamefont {Barry},\ and\ \citenamefont {DeMille}}]{Shuman:Nature467:820}%
  \BibitemOpen
  \bibfield  {author} {\bibinfo {author} {\bibfnamefont {E.~S.}\ \bibnamefont
  {Shuman}}, \bibinfo {author} {\bibfnamefont {J.~F.}\ \bibnamefont {Barry}}, \
  and\ \bibinfo {author} {\bibfnamefont {D.}~\bibnamefont {DeMille}},\
  }\bibfield  {title} {\enquote {\bibinfo {title} {Laser cooling of a diatomic
  molecule},}\ }\href {\doibase 10.1038/nature09443} {\bibfield  {journal}
  {\bibinfo  {journal} {Nature}\ }\textbf {\bibinfo {volume} {467}},\ \bibinfo
  {pages} {820--823} (\bibinfo {year} {2010})}\BibitemShut {NoStop}%
\bibitem [{\citenamefont {Zhelyazkova}\ \emph {et~al.}(2014)\citenamefont
  {Zhelyazkova}, \citenamefont {Cournol}, \citenamefont {Wall}, \citenamefont
  {Matsushima}, \citenamefont {Hudson}, \citenamefont {Hinds}, \citenamefont
  {Tarbutt},\ and\ \citenamefont {Sauer}}]{Zhelyazkova:PRA89:053416}%
  \BibitemOpen
  \bibfield  {author} {\bibinfo {author} {\bibfnamefont {V}~\bibnamefont
  {Zhelyazkova}}, \bibinfo {author} {\bibfnamefont {A}~\bibnamefont {Cournol}},
  \bibinfo {author} {\bibfnamefont {TE}~\bibnamefont {Wall}}, \bibinfo {author}
  {\bibfnamefont {A}~\bibnamefont {Matsushima}}, \bibinfo {author}
  {\bibfnamefont {JJ}~\bibnamefont {Hudson}}, \bibinfo {author} {\bibfnamefont
  {EA}~\bibnamefont {Hinds}}, \bibinfo {author} {\bibfnamefont
  {MR}~\bibnamefont {Tarbutt}}, \ and\ \bibinfo {author} {\bibfnamefont
  {BE}~\bibnamefont {Sauer}},\ }\bibfield  {title} {\enquote {\bibinfo {title}
  {Laser cooling and slowing of {CaF} molecules},}\ }\href {\doibase
  10.1103/PhysRevA.89.053416} {\bibfield  {journal} {\bibinfo  {journal}
  {Phys.\ Rev.\ A}\ }\textbf {\bibinfo {volume} {89}},\ \bibinfo {pages}
  {053416} (\bibinfo {year} {2014})}\BibitemShut {NoStop}%
\bibitem [{\citenamefont {Kozyryev}\ \emph {et~al.}(2017)\citenamefont
  {Kozyryev}, \citenamefont {Baum}, \citenamefont {Matsuda}, \citenamefont
  {Augenbraun}, \citenamefont {Anderegg}, \citenamefont {Sedlack},\ and\
  \citenamefont {Doyle}}]{Kozyryev:PRL118:173201}%
  \BibitemOpen
  \bibfield  {author} {\bibinfo {author} {\bibfnamefont {Ivan}\ \bibnamefont
  {Kozyryev}}, \bibinfo {author} {\bibfnamefont {Louis}\ \bibnamefont {Baum}},
  \bibinfo {author} {\bibfnamefont {Kyle}\ \bibnamefont {Matsuda}}, \bibinfo
  {author} {\bibfnamefont {Benjamin~L.}\ \bibnamefont {Augenbraun}}, \bibinfo
  {author} {\bibfnamefont {Loic}\ \bibnamefont {Anderegg}}, \bibinfo {author}
  {\bibfnamefont {Alexander~P.}\ \bibnamefont {Sedlack}}, \ and\ \bibinfo
  {author} {\bibfnamefont {John~M.}\ \bibnamefont {Doyle}},\ }\bibfield
  {title} {\enquote {\bibinfo {title} {Sisyphus laser cooling of a polyatomic
  molecule},}\ }\href {\doibase 10.1103/PhysRevLett.118.173201} {\bibfield
  {journal} {\bibinfo  {journal} {Phys.\ Rev.\ Lett.}\ }\textbf {\bibinfo
  {volume} {118}},\ \bibinfo {pages} {173201} (\bibinfo {year}
  {2017})}\BibitemShut {NoStop}%
\bibitem [{\citenamefont {Santamaria}\ \emph {et~al.}(2014)\citenamefont
  {Santamaria}, \citenamefont {Sarno}, \citenamefont {Ricciardi}, \citenamefont
  {Mosca}, \citenamefont {Rosa}, \citenamefont {Santambrogio}, \citenamefont
  {Maddaloni},\ and\ \citenamefont {Natale}}]{Santamaria:JMolSpec300:116}%
  \BibitemOpen
  \bibfield  {author} {\bibinfo {author} {\bibfnamefont {L.}~\bibnamefont
  {Santamaria}}, \bibinfo {author} {\bibfnamefont {V.~Di}\ \bibnamefont
  {Sarno}}, \bibinfo {author} {\bibfnamefont {I.}~\bibnamefont {Ricciardi}},
  \bibinfo {author} {\bibfnamefont {S.}~\bibnamefont {Mosca}}, \bibinfo
  {author} {\bibfnamefont {M.~De}\ \bibnamefont {Rosa}}, \bibinfo {author}
  {\bibfnamefont {G.}~\bibnamefont {Santambrogio}}, \bibinfo {author}
  {\bibfnamefont {P.}~\bibnamefont {Maddaloni}}, \ and\ \bibinfo {author}
  {\bibfnamefont {P.~De}\ \bibnamefont {Natale}},\ }\bibfield  {title}
  {\enquote {\bibinfo {title} {Assessing the time constancy of the
  proton-to-electron mass ratio by precision ro-vibrational spectroscopy of a
  cold molecular beam},}\ }\href {\doibase 10.1016/j.jms.2014.03.013}
  {\bibfield  {journal} {\bibinfo  {journal} {J.\ Mol.\ Spectrosc.}\ }\textbf
  {\bibinfo {volume} {300}},\ \bibinfo {pages} {116--123} (\bibinfo {year}
  {2014})}\BibitemShut {NoStop}%
\bibitem [{\citenamefont {Truppe}\ \emph {et~al.}(2013)\citenamefont {Truppe},
  \citenamefont {Hendricks}, \citenamefont {Tokunaga}, \citenamefont
  {Lewandowski}, \citenamefont {Kozlov}, \citenamefont {Henkel}, \citenamefont
  {Hinds},\ and\ \citenamefont {Tarbutt}}]{Truppe:NatComm4:2600}%
  \BibitemOpen
  \bibfield  {author} {\bibinfo {author} {\bibfnamefont {S.}~\bibnamefont
  {Truppe}}, \bibinfo {author} {\bibfnamefont {R.J.}\ \bibnamefont
  {Hendricks}}, \bibinfo {author} {\bibfnamefont {S.K.}\ \bibnamefont
  {Tokunaga}}, \bibinfo {author} {\bibfnamefont {H.J.}\ \bibnamefont
  {Lewandowski}}, \bibinfo {author} {\bibfnamefont {M.G.}\ \bibnamefont
  {Kozlov}}, \bibinfo {author} {\bibfnamefont {Christian}\ \bibnamefont
  {Henkel}}, \bibinfo {author} {\bibfnamefont {E.A.}\ \bibnamefont {Hinds}}, \
  and\ \bibinfo {author} {\bibfnamefont {M.R.}\ \bibnamefont {Tarbutt}},\
  }\bibfield  {title} {\enquote {\bibinfo {title} {A search for varying
  fundamental constants using hertz-level frequency measurements of cold {CH}
  molecules},}\ }\href {\doibase 10.1038/ncomms3600} {\bibfield  {journal}
  {\bibinfo  {journal} {Nat. Commun.}\ }\textbf {\bibinfo {volume} {4}},\
  \bibinfo {pages} {2600} (\bibinfo {year} {2013})}\BibitemShut {NoStop}%
\bibitem [{\citenamefont {Hudson}\ \emph {et~al.}(2011)\citenamefont {Hudson},
  \citenamefont {Kara}, \citenamefont {Smallman}, \citenamefont {Sauer},
  \citenamefont {Tarbutt},\ and\ \citenamefont {Hinds}}]{Hudson:Nature473:493}%
  \BibitemOpen
  \bibfield  {author} {\bibinfo {author} {\bibfnamefont {J.~J.}\ \bibnamefont
  {Hudson}}, \bibinfo {author} {\bibfnamefont {D.~M.}\ \bibnamefont {Kara}},
  \bibinfo {author} {\bibfnamefont {I.~J.}\ \bibnamefont {Smallman}}, \bibinfo
  {author} {\bibfnamefont {B.~E.}\ \bibnamefont {Sauer}}, \bibinfo {author}
  {\bibfnamefont {M.~R.}\ \bibnamefont {Tarbutt}}, \ and\ \bibinfo {author}
  {\bibfnamefont {E.~A.}\ \bibnamefont {Hinds}},\ }\bibfield  {title} {\enquote
  {\bibinfo {title} {Improved measurement of the shape of the electron},}\
  }\href {\doibase 10.1038/nature10104} {\bibfield  {journal} {\bibinfo
  {journal} {Nature}\ }\textbf {\bibinfo {volume} {473}},\ \bibinfo {pages}
  {493--496} (\bibinfo {year} {2011})}\BibitemShut {NoStop}%
\bibitem [{\citenamefont {Collaboration}\ \emph {et~al.}(2014)\citenamefont
  {Collaboration}, \citenamefont {Baron}, \citenamefont {Campbell},
  \citenamefont {DeMille}, \citenamefont {Doyle}, \citenamefont {Gabrielse},
  \citenamefont {Gurevich}, \citenamefont {Hess}, \citenamefont {Hutzler},
  \citenamefont {Kirilov}, \citenamefont {Kozyryev}, \citenamefont {O'Leary},
  \citenamefont {Panda}, \citenamefont {Petrik}, \citenamefont {Spaun},
  \citenamefont {Vutha},\ and\ \citenamefont {West}}]{ACME:Science343:269}%
  \BibitemOpen
  \bibfield  {author} {\bibinfo {author} {\bibfnamefont {The~ACME}\
  \bibnamefont {Collaboration}}, \bibinfo {author} {\bibfnamefont
  {J.}~\bibnamefont {Baron}}, \bibinfo {author} {\bibfnamefont {W.~C.}\
  \bibnamefont {Campbell}}, \bibinfo {author} {\bibfnamefont {D.}~\bibnamefont
  {DeMille}}, \bibinfo {author} {\bibfnamefont {J.~M.}\ \bibnamefont {Doyle}},
  \bibinfo {author} {\bibfnamefont {G.}~\bibnamefont {Gabrielse}}, \bibinfo
  {author} {\bibfnamefont {Y.~V.}\ \bibnamefont {Gurevich}}, \bibinfo {author}
  {\bibfnamefont {P.~W.}\ \bibnamefont {Hess}}, \bibinfo {author}
  {\bibfnamefont {N.~R.}\ \bibnamefont {Hutzler}}, \bibinfo {author}
  {\bibfnamefont {E.}~\bibnamefont {Kirilov}}, \bibinfo {author} {\bibfnamefont
  {I.}~\bibnamefont {Kozyryev}}, \bibinfo {author} {\bibfnamefont {B.~R.}\
  \bibnamefont {O'Leary}}, \bibinfo {author} {\bibfnamefont {C.~D.}\
  \bibnamefont {Panda}}, \bibinfo {author} {\bibfnamefont {E.~S.}\ \bibnamefont
  {Petrik}}, \bibinfo {author} {\bibfnamefont {B.}~\bibnamefont {Spaun}},
  \bibinfo {author} {\bibfnamefont {A.~C.}\ \bibnamefont {Vutha}}, \ and\
  \bibinfo {author} {\bibfnamefont {A.~D.}\ \bibnamefont {West}},\ }\bibfield
  {title} {\enquote {\bibinfo {title} {Order of magnitude smaller limit on the
  electric dipole moment of the electron},}\ }\href {\doibase
  10.1126/science.1248213} {\bibfield  {journal} {\bibinfo  {journal}
  {Science}\ }\textbf {\bibinfo {volume} {343}},\ \bibinfo {pages} {269--272}
  (\bibinfo {year} {2014})}\BibitemShut {NoStop}%
\bibitem [{\citenamefont {Seibert}\ \emph {et~al.}(2011)\citenamefont
  {Seibert}, \citenamefont {Ekeberg}, \citenamefont {Maia}, \citenamefont
  {Svenda}, \citenamefont {Andreasson}, \citenamefont {J{\"o}nsson},
  \citenamefont {Odi{\'c}}, \citenamefont {Iwan}, \citenamefont {Rocker},
  \citenamefont {Westphal}, \citenamefont {Hantke}, \citenamefont {Deponte},
  \citenamefont {Barty}, \citenamefont {Schulz}, \citenamefont {Gumprecht},
  \citenamefont {Coppola}, \citenamefont {Aquila}, \citenamefont {Liang},
  \citenamefont {White}, \citenamefont {Martin}, \citenamefont {Caleman},
  \citenamefont {Stern}, \citenamefont {Abergel}, \citenamefont {Seltzer},
  \citenamefont {Claverie}, \citenamefont {Bostedt}, \citenamefont {Bozek},
  \citenamefont {Boutet}, \citenamefont {Miahnahri}, \citenamefont
  {Messerschmidt}, \citenamefont {Krzywinski}, \citenamefont {Williams},
  \citenamefont {Hodgson}, \citenamefont {Bogan}, \citenamefont {Hampton},
  \citenamefont {Sierra}, \citenamefont {Starodub}, \citenamefont {Andersson},
  \citenamefont {Bajt}, \citenamefont {Barthelmess}, \citenamefont {Spence},
  \citenamefont {Fromme}, \citenamefont {Weierstall}, \citenamefont {Kirian},
  \citenamefont {Hunter}, \citenamefont {Doak}, \citenamefont {Marchesini},
  \citenamefont {Hau-Riege}, \citenamefont {Frank}, \citenamefont {Shoeman},
  \citenamefont {Lomb}, \citenamefont {Epp}, \citenamefont {Hartmann},
  \citenamefont {Rolles}, \citenamefont {Rudenko}, \citenamefont {Schmidt},
  \citenamefont {Foucar}, \citenamefont {Kimmel}, \citenamefont {Holl},
  \citenamefont {Rudek}, \citenamefont {Erk}, \citenamefont {H{\"o}mke},
  \citenamefont {Reich}, \citenamefont {Pietschner}, \citenamefont
  {Weidenspointner}, \citenamefont {Str{\"u}der}, \citenamefont {Hauser},
  \citenamefont {Gorke}, \citenamefont {Ullrich}, \citenamefont {Schlichting},
  \citenamefont {Herrmann}, \citenamefont {Schaller}, \citenamefont {Schopper},
  \citenamefont {Soltau}, \citenamefont {K{\"u}hnel}, \citenamefont
  {Andritschke}, \citenamefont {Schr{\"o}ter}, \citenamefont {Krasniqi},
  \citenamefont {Bott}, \citenamefont {Schorb}, \citenamefont {Rupp},
  \citenamefont {Adolph}, \citenamefont {Gorkhover}, \citenamefont {Hirsemann},
  \citenamefont {Potdevin}, \citenamefont {Graafsma}, \citenamefont {Nilsson},
  \citenamefont {Chapman},\ and\ \citenamefont {Hajdu}}]{Seibert:Nature470:78}%
  \BibitemOpen
  \bibfield  {author} {\bibinfo {author} {\bibfnamefont {M~Marvin}\
  \bibnamefont {Seibert}}, \bibinfo {author} {\bibfnamefont {Tomas}\
  \bibnamefont {Ekeberg}}, \bibinfo {author} {\bibfnamefont {Filipe R N~C}\
  \bibnamefont {Maia}}, \bibinfo {author} {\bibfnamefont {Martin}\ \bibnamefont
  {Svenda}}, \bibinfo {author} {\bibfnamefont {Jakob}\ \bibnamefont
  {Andreasson}}, \bibinfo {author} {\bibfnamefont {Olof}\ \bibnamefont
  {J{\"o}nsson}}, \bibinfo {author} {\bibfnamefont {Du{\v s}ko}\ \bibnamefont
  {Odi{\'c}}}, \bibinfo {author} {\bibfnamefont {Bianca}\ \bibnamefont {Iwan}},
  \bibinfo {author} {\bibfnamefont {Andrea}\ \bibnamefont {Rocker}}, \bibinfo
  {author} {\bibfnamefont {Daniel}\ \bibnamefont {Westphal}}, \bibinfo {author}
  {\bibfnamefont {Max}\ \bibnamefont {Hantke}}, \bibinfo {author}
  {\bibfnamefont {Daniel~P}\ \bibnamefont {Deponte}}, \bibinfo {author}
  {\bibfnamefont {Anton}\ \bibnamefont {Barty}}, \bibinfo {author}
  {\bibfnamefont {Joachim}\ \bibnamefont {Schulz}}, \bibinfo {author}
  {\bibfnamefont {Lars}\ \bibnamefont {Gumprecht}}, \bibinfo {author}
  {\bibfnamefont {Nicola}\ \bibnamefont {Coppola}}, \bibinfo {author}
  {\bibfnamefont {Andrew}\ \bibnamefont {Aquila}}, \bibinfo {author}
  {\bibfnamefont {Mengning}\ \bibnamefont {Liang}}, \bibinfo {author}
  {\bibfnamefont {Thomas~A}\ \bibnamefont {White}}, \bibinfo {author}
  {\bibfnamefont {Andrew}\ \bibnamefont {Martin}}, \bibinfo {author}
  {\bibfnamefont {Carl}\ \bibnamefont {Caleman}}, \bibinfo {author}
  {\bibfnamefont {Stephan}\ \bibnamefont {Stern}}, \bibinfo {author}
  {\bibfnamefont {Chantal}\ \bibnamefont {Abergel}}, \bibinfo {author}
  {\bibfnamefont {Virginie}\ \bibnamefont {Seltzer}}, \bibinfo {author}
  {\bibfnamefont {Jean-Michel}\ \bibnamefont {Claverie}}, \bibinfo {author}
  {\bibfnamefont {Christoph}\ \bibnamefont {Bostedt}}, \bibinfo {author}
  {\bibfnamefont {John~D}\ \bibnamefont {Bozek}}, \bibinfo {author}
  {\bibfnamefont {S{\'e}bastien}\ \bibnamefont {Boutet}}, \bibinfo {author}
  {\bibfnamefont {A~Alan}\ \bibnamefont {Miahnahri}}, \bibinfo {author}
  {\bibfnamefont {Marc}\ \bibnamefont {Messerschmidt}}, \bibinfo {author}
  {\bibfnamefont {Jacek}\ \bibnamefont {Krzywinski}}, \bibinfo {author}
  {\bibfnamefont {Garth}\ \bibnamefont {Williams}}, \bibinfo {author}
  {\bibfnamefont {Keith~O}\ \bibnamefont {Hodgson}}, \bibinfo {author}
  {\bibfnamefont {Michael~J}\ \bibnamefont {Bogan}}, \bibinfo {author}
  {\bibfnamefont {Christina~Y}\ \bibnamefont {Hampton}}, \bibinfo {author}
  {\bibfnamefont {Raymond~G}\ \bibnamefont {Sierra}}, \bibinfo {author}
  {\bibfnamefont {Dmitri}\ \bibnamefont {Starodub}}, \bibinfo {author}
  {\bibfnamefont {Inger}\ \bibnamefont {Andersson}}, \bibinfo {author}
  {\bibfnamefont {Sa{\v s}a}\ \bibnamefont {Bajt}}, \bibinfo {author}
  {\bibfnamefont {Miriam}\ \bibnamefont {Barthelmess}}, \bibinfo {author}
  {\bibfnamefont {John C~H}\ \bibnamefont {Spence}}, \bibinfo {author}
  {\bibfnamefont {Petra}\ \bibnamefont {Fromme}}, \bibinfo {author}
  {\bibfnamefont {Uwe}\ \bibnamefont {Weierstall}}, \bibinfo {author}
  {\bibfnamefont {Richard}\ \bibnamefont {Kirian}}, \bibinfo {author}
  {\bibfnamefont {Mark}\ \bibnamefont {Hunter}}, \bibinfo {author}
  {\bibfnamefont {R~Bruce}\ \bibnamefont {Doak}}, \bibinfo {author}
  {\bibfnamefont {Stefano}\ \bibnamefont {Marchesini}}, \bibinfo {author}
  {\bibfnamefont {Stefan~P}\ \bibnamefont {Hau-Riege}}, \bibinfo {author}
  {\bibfnamefont {Matthias}\ \bibnamefont {Frank}}, \bibinfo {author}
  {\bibfnamefont {Robert~L}\ \bibnamefont {Shoeman}}, \bibinfo {author}
  {\bibfnamefont {Lukas}\ \bibnamefont {Lomb}}, \bibinfo {author}
  {\bibfnamefont {Sascha~W}\ \bibnamefont {Epp}}, \bibinfo {author}
  {\bibfnamefont {Robert}\ \bibnamefont {Hartmann}}, \bibinfo {author}
  {\bibfnamefont {Daniel}\ \bibnamefont {Rolles}}, \bibinfo {author}
  {\bibfnamefont {Artem}\ \bibnamefont {Rudenko}}, \bibinfo {author}
  {\bibfnamefont {Carlo}\ \bibnamefont {Schmidt}}, \bibinfo {author}
  {\bibfnamefont {Lutz}\ \bibnamefont {Foucar}}, \bibinfo {author}
  {\bibfnamefont {Nils}\ \bibnamefont {Kimmel}}, \bibinfo {author}
  {\bibfnamefont {Peter}\ \bibnamefont {Holl}}, \bibinfo {author}
  {\bibfnamefont {Benedikt}\ \bibnamefont {Rudek}}, \bibinfo {author}
  {\bibfnamefont {Benjamin}\ \bibnamefont {Erk}}, \bibinfo {author}
  {\bibfnamefont {Andr{\'e}}\ \bibnamefont {H{\"o}mke}}, \bibinfo {author}
  {\bibfnamefont {Christian}\ \bibnamefont {Reich}}, \bibinfo {author}
  {\bibfnamefont {Daniel}\ \bibnamefont {Pietschner}}, \bibinfo {author}
  {\bibfnamefont {Georg}\ \bibnamefont {Weidenspointner}}, \bibinfo {author}
  {\bibfnamefont {Lothar}\ \bibnamefont {Str{\"u}der}}, \bibinfo {author}
  {\bibfnamefont {G{\"u}nter}\ \bibnamefont {Hauser}}, \bibinfo {author}
  {\bibfnamefont {Hubert}\ \bibnamefont {Gorke}}, \bibinfo {author}
  {\bibfnamefont {Joachim}\ \bibnamefont {Ullrich}}, \bibinfo {author}
  {\bibfnamefont {Ilme}\ \bibnamefont {Schlichting}}, \bibinfo {author}
  {\bibfnamefont {Sven}\ \bibnamefont {Herrmann}}, \bibinfo {author}
  {\bibfnamefont {Gerhard}\ \bibnamefont {Schaller}}, \bibinfo {author}
  {\bibfnamefont {Florian}\ \bibnamefont {Schopper}}, \bibinfo {author}
  {\bibfnamefont {Heike}\ \bibnamefont {Soltau}}, \bibinfo {author}
  {\bibfnamefont {Kai-Uwe}\ \bibnamefont {K{\"u}hnel}}, \bibinfo {author}
  {\bibfnamefont {Robert}\ \bibnamefont {Andritschke}}, \bibinfo {author}
  {\bibfnamefont {Claus-Dieter}\ \bibnamefont {Schr{\"o}ter}}, \bibinfo
  {author} {\bibfnamefont {Faton}\ \bibnamefont {Krasniqi}}, \bibinfo {author}
  {\bibfnamefont {Mario}\ \bibnamefont {Bott}}, \bibinfo {author}
  {\bibfnamefont {Sebastian}\ \bibnamefont {Schorb}}, \bibinfo {author}
  {\bibfnamefont {Daniela}\ \bibnamefont {Rupp}}, \bibinfo {author}
  {\bibfnamefont {Marcus}\ \bibnamefont {Adolph}}, \bibinfo {author}
  {\bibfnamefont {Tais}\ \bibnamefont {Gorkhover}}, \bibinfo {author}
  {\bibfnamefont {Helmut}\ \bibnamefont {Hirsemann}}, \bibinfo {author}
  {\bibfnamefont {Guillaume}\ \bibnamefont {Potdevin}}, \bibinfo {author}
  {\bibfnamefont {Heinz}\ \bibnamefont {Graafsma}}, \bibinfo {author}
  {\bibfnamefont {Bj{\"o}rn}\ \bibnamefont {Nilsson}}, \bibinfo {author}
  {\bibfnamefont {Henry~N}\ \bibnamefont {Chapman}}, \ and\ \bibinfo {author}
  {\bibfnamefont {Janos}\ \bibnamefont {Hajdu}},\ }\bibfield  {title} {\enquote
  {\bibinfo {title} {Single mimivirus particles intercepted and imaged with an
  x-ray laser},}\ }\href {\doibase 10.1038/nature09748} {\bibfield  {journal}
  {\bibinfo  {journal} {Nature}\ }\textbf {\bibinfo {volume} {470}},\ \bibinfo
  {pages} {78} (\bibinfo {year} {2011})}\BibitemShut {NoStop}%
\bibitem [{\citenamefont {Ekeberg}\ \emph {et~al.}(2015)\citenamefont
  {Ekeberg}, \citenamefont {Svenda}, \citenamefont {Abergel}, \citenamefont
  {Maia}, \citenamefont {Seltzer}, \citenamefont {Claverie}, \citenamefont
  {Hantke}, \citenamefont {J{\"o}nsson}, \citenamefont {Nettelblad},
  \citenamefont {van~der Schot}, \citenamefont {Liang}, \citenamefont
  {Deponte}, \citenamefont {Barty}, \citenamefont {Seibert}, \citenamefont
  {Iwan}, \citenamefont {Andersson}, \citenamefont {Loh}, \citenamefont
  {Martin}, \citenamefont {Chapman}, \citenamefont {Bostedt}, \citenamefont
  {Bozek}, \citenamefont {Ferguson}, \citenamefont {Krzywinski}, \citenamefont
  {Epp}, \citenamefont {Rolles}, \citenamefont {Rudenko}, \citenamefont
  {Hartmann}, \citenamefont {Kimmel},\ and\ \citenamefont
  {Hajdu}}]{Ekeberg:PRL114:098102}%
  \BibitemOpen
  \bibfield  {author} {\bibinfo {author} {\bibfnamefont {Tomas}\ \bibnamefont
  {Ekeberg}}, \bibinfo {author} {\bibfnamefont {Martin}\ \bibnamefont
  {Svenda}}, \bibinfo {author} {\bibfnamefont {Chantal}\ \bibnamefont
  {Abergel}}, \bibinfo {author} {\bibfnamefont {Filipe R N~C}\ \bibnamefont
  {Maia}}, \bibinfo {author} {\bibfnamefont {Virginie}\ \bibnamefont
  {Seltzer}}, \bibinfo {author} {\bibfnamefont {Jean-Michel}\ \bibnamefont
  {Claverie}}, \bibinfo {author} {\bibfnamefont {Max}\ \bibnamefont {Hantke}},
  \bibinfo {author} {\bibfnamefont {Olof}\ \bibnamefont {J{\"o}nsson}},
  \bibinfo {author} {\bibfnamefont {Carl}\ \bibnamefont {Nettelblad}}, \bibinfo
  {author} {\bibfnamefont {Gijs}\ \bibnamefont {van~der Schot}}, \bibinfo
  {author} {\bibfnamefont {Mengning}\ \bibnamefont {Liang}}, \bibinfo {author}
  {\bibfnamefont {Daniel~P}\ \bibnamefont {Deponte}}, \bibinfo {author}
  {\bibfnamefont {Anton}\ \bibnamefont {Barty}}, \bibinfo {author}
  {\bibfnamefont {M~Marvin}\ \bibnamefont {Seibert}}, \bibinfo {author}
  {\bibfnamefont {Bianca}\ \bibnamefont {Iwan}}, \bibinfo {author}
  {\bibfnamefont {Inger}\ \bibnamefont {Andersson}}, \bibinfo {author}
  {\bibfnamefont {N~Duane}\ \bibnamefont {Loh}}, \bibinfo {author}
  {\bibfnamefont {Andrew~V}\ \bibnamefont {Martin}}, \bibinfo {author}
  {\bibfnamefont {Henry}\ \bibnamefont {Chapman}}, \bibinfo {author}
  {\bibfnamefont {Christoph}\ \bibnamefont {Bostedt}}, \bibinfo {author}
  {\bibfnamefont {John~D}\ \bibnamefont {Bozek}}, \bibinfo {author}
  {\bibfnamefont {Ken~R}\ \bibnamefont {Ferguson}}, \bibinfo {author}
  {\bibfnamefont {Jacek}\ \bibnamefont {Krzywinski}}, \bibinfo {author}
  {\bibfnamefont {Sascha~W}\ \bibnamefont {Epp}}, \bibinfo {author}
  {\bibfnamefont {Daniel}\ \bibnamefont {Rolles}}, \bibinfo {author}
  {\bibfnamefont {Artem}\ \bibnamefont {Rudenko}}, \bibinfo {author}
  {\bibfnamefont {Robert}\ \bibnamefont {Hartmann}}, \bibinfo {author}
  {\bibfnamefont {Nils}\ \bibnamefont {Kimmel}}, \ and\ \bibinfo {author}
  {\bibfnamefont {Janos}\ \bibnamefont {Hajdu}},\ }\bibfield  {title} {\enquote
  {\bibinfo {title} {Three-dimensional reconstruction of the giant mimivirus
  particle with an x-ray free-electron laser},}\ }\href {\doibase
  https://doi.org/10.1103/PhysRevLett.114.098102} {\bibfield  {journal}
  {\bibinfo  {journal} {Phys.\ Rev.\ Lett.}\ }\textbf {\bibinfo {volume}
  {114}},\ \bibinfo {pages} {098102} (\bibinfo {year} {2015})}\BibitemShut
  {NoStop}%
\bibitem [{\citenamefont {Barty}\ \emph {et~al.}(2013)\citenamefont {Barty},
  \citenamefont {K{\"u}pper},\ and\ \citenamefont
  {Chapman}}]{Barty:ARPC64:415}%
  \BibitemOpen
  \bibfield  {author} {\bibinfo {author} {\bibfnamefont {Anton}\ \bibnamefont
  {Barty}}, \bibinfo {author} {\bibfnamefont {Jochen}\ \bibnamefont
  {K{\"u}pper}}, \ and\ \bibinfo {author} {\bibfnamefont {Henry~N.}\
  \bibnamefont {Chapman}},\ }\bibfield  {title} {\enquote {\bibinfo {title}
  {Molecular imaging using x-ray free-electron lasers},}\ }\href {\doibase
  10.1146/annurev-physchem-032511-143708} {\bibfield  {journal} {\bibinfo
  {journal} {Annu.\ Rev.\ Phys.\ Chem.}\ }\textbf {\bibinfo {volume} {64}},\
  \bibinfo {pages} {415--435} (\bibinfo {year} {2013})}\BibitemShut {NoStop}%
\bibitem [{\citenamefont {K{\"u}pper}\ \emph {et~al.}(2014)\citenamefont
  {K{\"u}pper}, \citenamefont {Stern}, \citenamefont {Holmegaard},
  \citenamefont {Filsinger}, \citenamefont {Rouz\'{e}e}, \citenamefont
  {Rudenko}, \citenamefont {Johnsson}, \citenamefont {Martin}, \citenamefont
  {Adolph}, \citenamefont {Aquila}, \citenamefont {Bajt}, \citenamefont
  {Barty}, \citenamefont {Bostedt}, \citenamefont {Bozek}, \citenamefont
  {Caleman}, \citenamefont {Coffee}, \citenamefont {Coppola}, \citenamefont
  {Delmas}, \citenamefont {Epp}, \citenamefont {Erk}, \citenamefont {Foucar},
  \citenamefont {Gorkhover}, \citenamefont {Gumprecht}, \citenamefont
  {Hartmann}, \citenamefont {Hartmann}, \citenamefont {Hauser}, \citenamefont
  {Holl}, \citenamefont {H{\"o}mke}, \citenamefont {Kimmel}, \citenamefont
  {Krasniqi}, \citenamefont {K{\"u}hnel}, \citenamefont {Maurer}, \citenamefont
  {Messerschmidt}, \citenamefont {Moshammer}, \citenamefont {Reich},
  \citenamefont {Rudek}, \citenamefont {Santra}, \citenamefont {Schlichting},
  \citenamefont {Schmidt}, \citenamefont {Schorb}, \citenamefont {Schulz},
  \citenamefont {Soltau}, \citenamefont {Spence}, \citenamefont {Starodub},
  \citenamefont {Str{\"u}der}, \citenamefont {Th{\o}gersen}, \citenamefont
  {Vrakking}, \citenamefont {Weidenspointner}, \citenamefont {White},
  \citenamefont {Wunderer}, \citenamefont {Meijer}, \citenamefont {Ullrich},
  \citenamefont {Stapelfeldt}, \citenamefont {Rolles},\ and\ \citenamefont
  {Chapman}}]{Kuepper:PRL112:083002}%
  \BibitemOpen
  \bibfield  {author} {\bibinfo {author} {\bibfnamefont {Jochen}\ \bibnamefont
  {K{\"u}pper}}, \bibinfo {author} {\bibfnamefont {Stephan}\ \bibnamefont
  {Stern}}, \bibinfo {author} {\bibfnamefont {Lotte}\ \bibnamefont
  {Holmegaard}}, \bibinfo {author} {\bibfnamefont {Frank}\ \bibnamefont
  {Filsinger}}, \bibinfo {author} {\bibfnamefont {Arnaud}\ \bibnamefont
  {Rouz\'{e}e}}, \bibinfo {author} {\bibfnamefont {Artem}\ \bibnamefont
  {Rudenko}}, \bibinfo {author} {\bibfnamefont {Per}\ \bibnamefont {Johnsson}},
  \bibinfo {author} {\bibfnamefont {Andrew~V.}\ \bibnamefont {Martin}},
  \bibinfo {author} {\bibfnamefont {Marcus}\ \bibnamefont {Adolph}}, \bibinfo
  {author} {\bibfnamefont {Andrew}\ \bibnamefont {Aquila}}, \bibinfo {author}
  {\bibfnamefont {Sa{\v s}a}\ \bibnamefont {Bajt}}, \bibinfo {author}
  {\bibfnamefont {Anton}\ \bibnamefont {Barty}}, \bibinfo {author}
  {\bibfnamefont {Christoph}\ \bibnamefont {Bostedt}}, \bibinfo {author}
  {\bibfnamefont {John}\ \bibnamefont {Bozek}}, \bibinfo {author}
  {\bibfnamefont {Carl}\ \bibnamefont {Caleman}}, \bibinfo {author}
  {\bibfnamefont {Ryan}\ \bibnamefont {Coffee}}, \bibinfo {author}
  {\bibfnamefont {Nicola}\ \bibnamefont {Coppola}}, \bibinfo {author}
  {\bibfnamefont {Tjark}\ \bibnamefont {Delmas}}, \bibinfo {author}
  {\bibfnamefont {Sascha}\ \bibnamefont {Epp}}, \bibinfo {author}
  {\bibfnamefont {Benjamin}\ \bibnamefont {Erk}}, \bibinfo {author}
  {\bibfnamefont {Lutz}\ \bibnamefont {Foucar}}, \bibinfo {author}
  {\bibfnamefont {Tais}\ \bibnamefont {Gorkhover}}, \bibinfo {author}
  {\bibfnamefont {Lars}\ \bibnamefont {Gumprecht}}, \bibinfo {author}
  {\bibfnamefont {Andreas}\ \bibnamefont {Hartmann}}, \bibinfo {author}
  {\bibfnamefont {Robert}\ \bibnamefont {Hartmann}}, \bibinfo {author}
  {\bibfnamefont {G{\"u}nter}\ \bibnamefont {Hauser}}, \bibinfo {author}
  {\bibfnamefont {Peter}\ \bibnamefont {Holl}}, \bibinfo {author}
  {\bibfnamefont {Andre}\ \bibnamefont {H{\"o}mke}}, \bibinfo {author}
  {\bibfnamefont {Nils}\ \bibnamefont {Kimmel}}, \bibinfo {author}
  {\bibfnamefont {Faton}\ \bibnamefont {Krasniqi}}, \bibinfo {author}
  {\bibfnamefont {Kai-Uwe}\ \bibnamefont {K{\"u}hnel}}, \bibinfo {author}
  {\bibfnamefont {Jochen}\ \bibnamefont {Maurer}}, \bibinfo {author}
  {\bibfnamefont {Marc}\ \bibnamefont {Messerschmidt}}, \bibinfo {author}
  {\bibfnamefont {Robert}\ \bibnamefont {Moshammer}}, \bibinfo {author}
  {\bibfnamefont {Christian}\ \bibnamefont {Reich}}, \bibinfo {author}
  {\bibfnamefont {Benedikt}\ \bibnamefont {Rudek}}, \bibinfo {author}
  {\bibfnamefont {Robin}\ \bibnamefont {Santra}}, \bibinfo {author}
  {\bibfnamefont {Ilme}\ \bibnamefont {Schlichting}}, \bibinfo {author}
  {\bibfnamefont {Carlo}\ \bibnamefont {Schmidt}}, \bibinfo {author}
  {\bibfnamefont {Sebastian}\ \bibnamefont {Schorb}}, \bibinfo {author}
  {\bibfnamefont {Joachim}\ \bibnamefont {Schulz}}, \bibinfo {author}
  {\bibfnamefont {Heike}\ \bibnamefont {Soltau}}, \bibinfo {author}
  {\bibfnamefont {John C.~H.}\ \bibnamefont {Spence}}, \bibinfo {author}
  {\bibfnamefont {Dmitri}\ \bibnamefont {Starodub}}, \bibinfo {author}
  {\bibfnamefont {Lothar}\ \bibnamefont {Str{\"u}der}}, \bibinfo {author}
  {\bibfnamefont {Jan}\ \bibnamefont {Th{\o}gersen}}, \bibinfo {author}
  {\bibfnamefont {Marc J.~J.}\ \bibnamefont {Vrakking}}, \bibinfo {author}
  {\bibfnamefont {Georg}\ \bibnamefont {Weidenspointner}}, \bibinfo {author}
  {\bibfnamefont {Thomas~A.}\ \bibnamefont {White}}, \bibinfo {author}
  {\bibfnamefont {Cornelia}\ \bibnamefont {Wunderer}}, \bibinfo {author}
  {\bibfnamefont {Gerard}\ \bibnamefont {Meijer}}, \bibinfo {author}
  {\bibfnamefont {Joachim}\ \bibnamefont {Ullrich}}, \bibinfo {author}
  {\bibfnamefont {Henrik}\ \bibnamefont {Stapelfeldt}}, \bibinfo {author}
  {\bibfnamefont {Daniel}\ \bibnamefont {Rolles}}, \ and\ \bibinfo {author}
  {\bibfnamefont {Henry~N.}\ \bibnamefont {Chapman}},\ }\bibfield  {title}
  {\enquote {\bibinfo {title} {X-ray diffraction from isolated and strongly
  aligned gas-phase molecules with a free-electron laser},}\ }\href {\doibase
  10.1103/PhysRevLett.112.083002} {\bibfield  {journal} {\bibinfo  {journal}
  {Phys.\ Rev.\ Lett.}\ }\textbf {\bibinfo {volume} {112}},\ \bibinfo {pages}
  {083002} (\bibinfo {year} {2014})},\ \Eprint {http://arxiv.org/abs/1307.4577}
  {arXiv:1307.4577 [physics]} \BibitemShut {NoStop}%
\bibitem [{\citenamefont {Filsinger}\ \emph {et~al.}(2011)\citenamefont
  {Filsinger}, \citenamefont {Meijer}, \citenamefont {Stapelfeldt},
  \citenamefont {Chapman},\ and\ \citenamefont
  {K{\"u}pper}}]{Filsinger:PCCP13:2076}%
  \BibitemOpen
  \bibfield  {author} {\bibinfo {author} {\bibfnamefont {Frank}\ \bibnamefont
  {Filsinger}}, \bibinfo {author} {\bibfnamefont {Gerard}\ \bibnamefont
  {Meijer}}, \bibinfo {author} {\bibfnamefont {Henrik}\ \bibnamefont
  {Stapelfeldt}}, \bibinfo {author} {\bibfnamefont {Henry}\ \bibnamefont
  {Chapman}}, \ and\ \bibinfo {author} {\bibfnamefont {Jochen}\ \bibnamefont
  {K{\"u}pper}},\ }\bibfield  {title} {\enquote {\bibinfo {title} {State- and
  conformer-selected beams of aligned and oriented molecules for ultrafast
  diffraction studies},}\ }\href {\doibase 10.1039/C0CP01585G} {\bibfield
  {journal} {\bibinfo  {journal} {Phys.\ Chem.\ Chem.\ Phys.}\ }\textbf
  {\bibinfo {volume} {13}},\ \bibinfo {pages} {2076--2087} (\bibinfo {year}
  {2011})},\ \Eprint {http://arxiv.org/abs/1009.0871} {arXiv:1009.0871
  [physics]} \BibitemShut {NoStop}%
\bibitem [{\citenamefont {Filsinger}\ \emph {et~al.}(2008)\citenamefont
  {Filsinger}, \citenamefont {Erlekam}, \citenamefont {von Helden},
  \citenamefont {K{\"u}pper},\ and\ \citenamefont
  {Meijer}}]{Filsinger:PRL100:133003}%
  \BibitemOpen
  \bibfield  {author} {\bibinfo {author} {\bibfnamefont {Frank}\ \bibnamefont
  {Filsinger}}, \bibinfo {author} {\bibfnamefont {Undine}\ \bibnamefont
  {Erlekam}}, \bibinfo {author} {\bibfnamefont {Gert}\ \bibnamefont {von
  Helden}}, \bibinfo {author} {\bibfnamefont {Jochen}\ \bibnamefont
  {K{\"u}pper}}, \ and\ \bibinfo {author} {\bibfnamefont {Gerard}\ \bibnamefont
  {Meijer}},\ }\bibfield  {title} {\enquote {\bibinfo {title} {Selector for
  structural isomers of neutral molecules},}\ }\href {\doibase
  10.1103/PhysRevLett.100.133003} {\bibfield  {journal} {\bibinfo  {journal}
  {Phys.\ Rev.\ Lett.}\ }\textbf {\bibinfo {volume} {100}},\ \bibinfo {pages}
  {133003} (\bibinfo {year} {2008})},\ \Eprint {http://arxiv.org/abs/0802.2795}
  {arXiv:0802.2795 [physics]} \BibitemShut {NoStop}%
\bibitem [{\citenamefont {Chang}\ \emph {et~al.}(2015)\citenamefont {Chang},
  \citenamefont {Horke}, \citenamefont {Trippel},\ and\ \citenamefont
  {K{\"u}pper}}]{Chang:IRPC34:557}%
  \BibitemOpen
  \bibfield  {author} {\bibinfo {author} {\bibfnamefont {Yuan-Pin}\
  \bibnamefont {Chang}}, \bibinfo {author} {\bibfnamefont {Daniel~A.}\
  \bibnamefont {Horke}}, \bibinfo {author} {\bibfnamefont {Sebastian}\
  \bibnamefont {Trippel}}, \ and\ \bibinfo {author} {\bibfnamefont {Jochen}\
  \bibnamefont {K{\"u}pper}},\ }\bibfield  {title} {\enquote {\bibinfo {title}
  {Spatially-controlled complex molecules and their applications},}\ }\href
  {\doibase 10.1080/0144235X.2015.1077838} {\bibfield  {journal} {\bibinfo
  {journal} {Int.\ Rev.\ Phys.\ Chem.}\ }\textbf {\bibinfo {volume} {34}},\
  \bibinfo {pages} {557--590} (\bibinfo {year} {2015})},\ \Eprint
  {http://arxiv.org/abs/1505.05632} {arXiv:1505.05632 [physics]} \BibitemShut
  {NoStop}%
\bibitem [{\citenamefont {Holmegaard}\ \emph {et~al.}(2009)\citenamefont
  {Holmegaard}, \citenamefont {Nielsen}, \citenamefont {Nevo}, \citenamefont
  {Stapelfeldt}, \citenamefont {Filsinger}, \citenamefont {K{\"u}pper},\ and\
  \citenamefont {Meijer}}]{Holmegaard:PRL102:023001}%
  \BibitemOpen
  \bibfield  {author} {\bibinfo {author} {\bibfnamefont {Lotte}\ \bibnamefont
  {Holmegaard}}, \bibinfo {author} {\bibfnamefont {Jens~H.}\ \bibnamefont
  {Nielsen}}, \bibinfo {author} {\bibfnamefont {Iftach}\ \bibnamefont {Nevo}},
  \bibinfo {author} {\bibfnamefont {Henrik}\ \bibnamefont {Stapelfeldt}},
  \bibinfo {author} {\bibfnamefont {Frank}\ \bibnamefont {Filsinger}}, \bibinfo
  {author} {\bibfnamefont {Jochen}\ \bibnamefont {K{\"u}pper}}, \ and\ \bibinfo
  {author} {\bibfnamefont {Gerard}\ \bibnamefont {Meijer}},\ }\bibfield
  {title} {\enquote {\bibinfo {title} {Laser-induced alignment and orientation
  of quantum-state-selected large molecules},}\ }\href {\doibase
  10.1103/PhysRevLett.102.023001} {\bibfield  {journal} {\bibinfo  {journal}
  {Phys.\ Rev.\ Lett.}\ }\textbf {\bibinfo {volume} {102}},\ \bibinfo {pages}
  {023001} (\bibinfo {year} {2009})},\ \Eprint {http://arxiv.org/abs/0810.2307}
  {arXiv:0810.2307 [physics]} \BibitemShut {NoStop}%
\bibitem [{\citenamefont {Filsinger}\ \emph {et~al.}(2009)\citenamefont
  {Filsinger}, \citenamefont {K{\"u}pper}, \citenamefont {Meijer},
  \citenamefont {Holmegaard}, \citenamefont {Nielsen}, \citenamefont {Nevo},
  \citenamefont {Hansen},\ and\ \citenamefont
  {Stapelfeldt}}]{Filsinger:JCP131:064309}%
  \BibitemOpen
  \bibfield  {author} {\bibinfo {author} {\bibfnamefont {Frank}\ \bibnamefont
  {Filsinger}}, \bibinfo {author} {\bibfnamefont {Jochen}\ \bibnamefont
  {K{\"u}pper}}, \bibinfo {author} {\bibfnamefont {Gerard}\ \bibnamefont
  {Meijer}}, \bibinfo {author} {\bibfnamefont {Lotte}\ \bibnamefont
  {Holmegaard}}, \bibinfo {author} {\bibfnamefont {Jens~H.}\ \bibnamefont
  {Nielsen}}, \bibinfo {author} {\bibfnamefont {Iftach}\ \bibnamefont {Nevo}},
  \bibinfo {author} {\bibfnamefont {Jonas~L.}\ \bibnamefont {Hansen}}, \ and\
  \bibinfo {author} {\bibfnamefont {Henrik}\ \bibnamefont {Stapelfeldt}},\
  }\bibfield  {title} {\enquote {\bibinfo {title} {Quantum-state selection,
  alignment, and orientation of large molecules using static electric and laser
  fields},}\ }\href {\doibase 10.1063/1.3194287} {\bibfield  {journal}
  {\bibinfo  {journal} {J.\ Chem.\ Phys.}\ }\textbf {\bibinfo {volume} {131}},\
  \bibinfo {pages} {064309} (\bibinfo {year} {2009})},\ \Eprint
  {http://arxiv.org/abs/0903.5413} {arXiv:0903.5413 [physics]} \BibitemShut
  {NoStop}%
\bibitem [{\citenamefont {Nevo}\ \emph {et~al.}(2009)\citenamefont {Nevo},
  \citenamefont {Holmegaard}, \citenamefont {Nielsen}, \citenamefont {Hansen},
  \citenamefont {Stapelfeldt}, \citenamefont {Filsinger}, \citenamefont
  {Meijer},\ and\ \citenamefont {K{\"u}pper}}]{Nevo:PCCP11:9912}%
  \BibitemOpen
  \bibfield  {author} {\bibinfo {author} {\bibfnamefont {Iftach}\ \bibnamefont
  {Nevo}}, \bibinfo {author} {\bibfnamefont {Lotte}\ \bibnamefont
  {Holmegaard}}, \bibinfo {author} {\bibfnamefont {Jens~H.}\ \bibnamefont
  {Nielsen}}, \bibinfo {author} {\bibfnamefont {Jonas~L.}\ \bibnamefont
  {Hansen}}, \bibinfo {author} {\bibfnamefont {Henrik}\ \bibnamefont
  {Stapelfeldt}}, \bibinfo {author} {\bibfnamefont {Frank}\ \bibnamefont
  {Filsinger}}, \bibinfo {author} {\bibfnamefont {Gerard}\ \bibnamefont
  {Meijer}}, \ and\ \bibinfo {author} {\bibfnamefont {Jochen}\ \bibnamefont
  {K{\"u}pper}},\ }\bibfield  {title} {\enquote {\bibinfo {title}
  {Laser-induced 3{D} alignment and orientation of quantum state-selected
  molecules},}\ }\href {\doibase 10.1039/b910423b} {\bibfield  {journal}
  {\bibinfo  {journal} {Phys.\ Chem.\ Chem.\ Phys.}\ }\textbf {\bibinfo
  {volume} {11}},\ \bibinfo {pages} {9912--9918} (\bibinfo {year} {2009})},\
  \Eprint {http://arxiv.org/abs/0906.2971} {arXiv:0906.2971 [physics]}
  \BibitemShut {NoStop}%
\bibitem [{\citenamefont {Trippel}\ \emph {et~al.}(2013)\citenamefont
  {Trippel}, \citenamefont {Mullins}, \citenamefont {M{\"u}ller}, \citenamefont
  {Kienitz}, \citenamefont {D{\l}ugo{\l}\k{e}cki},\ and\ \citenamefont
  {K{\"u}pper}}]{Trippel:MP111:1738}%
  \BibitemOpen
  \bibfield  {author} {\bibinfo {author} {\bibfnamefont {Sebastian}\
  \bibnamefont {Trippel}}, \bibinfo {author} {\bibfnamefont {Terry}\
  \bibnamefont {Mullins}}, \bibinfo {author} {\bibfnamefont {Nele L.~M.}\
  \bibnamefont {M{\"u}ller}}, \bibinfo {author} {\bibfnamefont {Jens~S.}\
  \bibnamefont {Kienitz}}, \bibinfo {author} {\bibfnamefont {Karol}\
  \bibnamefont {D{\l}ugo{\l}\k{e}cki}}, \ and\ \bibinfo {author} {\bibfnamefont
  {Jochen}\ \bibnamefont {K{\"u}pper}},\ }\bibfield  {title} {\enquote
  {\bibinfo {title} {Strongly aligned and oriented molecular samples at a {kHz}
  repetition rate},}\ }\href {\doibase 10.1080/00268976.2013.780334} {\bibfield
   {journal} {\bibinfo  {journal} {Mol.\ Phys.}\ }\textbf {\bibinfo {volume}
  {111}},\ \bibinfo {pages} {1738} (\bibinfo {year} {2013})},\ \Eprint
  {http://arxiv.org/abs/1301.1826} {arXiv:1301.1826 [physics]} \BibitemShut
  {NoStop}%
\bibitem [{\citenamefont {Trippel}\ \emph {et~al.}(2014)\citenamefont
  {Trippel}, \citenamefont {Mullins}, \citenamefont {M{\"u}ller}, \citenamefont
  {Kienitz}, \citenamefont {Omiste}, \citenamefont {Stapelfeldt}, \citenamefont
  {Gonz{\'a}lez-F{\'e}rez},\ and\ \citenamefont
  {K{\"u}pper}}]{Trippel:PRA89:051401R}%
  \BibitemOpen
  \bibfield  {author} {\bibinfo {author} {\bibfnamefont {Sebastian}\
  \bibnamefont {Trippel}}, \bibinfo {author} {\bibfnamefont {Terence}\
  \bibnamefont {Mullins}}, \bibinfo {author} {\bibfnamefont {N~L~M}\
  \bibnamefont {M{\"u}ller}}, \bibinfo {author} {\bibfnamefont {Jens~S}\
  \bibnamefont {Kienitz}}, \bibinfo {author} {\bibfnamefont {Juan~J}\
  \bibnamefont {Omiste}}, \bibinfo {author} {\bibfnamefont {Henrik}\
  \bibnamefont {Stapelfeldt}}, \bibinfo {author} {\bibfnamefont {Rosario}\
  \bibnamefont {Gonz{\'a}lez-F{\'e}rez}}, \ and\ \bibinfo {author}
  {\bibfnamefont {Jochen}\ \bibnamefont {K{\"u}pper}},\ }\bibfield  {title}
  {\enquote {\bibinfo {title} {Strongly driven quantum pendulum of the carbonyl
  sulfide molecule},}\ }\href {\doibase 10.1103/PhysRevA.89.051401} {\bibfield
  {journal} {\bibinfo  {journal} {Phys.\ Rev.\ A}\ }\textbf {\bibinfo {volume}
  {89}},\ \bibinfo {pages} {051401(R)} (\bibinfo {year} {2014})},\ \Eprint
  {http://arxiv.org/abs/1401.6897} {arXiv:1401.6897 [quant-ph]} \BibitemShut
  {NoStop}%
\bibitem [{\citenamefont {Trippel}\ \emph {et~al.}(2015)\citenamefont
  {Trippel}, \citenamefont {Mullins}, \citenamefont {M{\"u}ller}, \citenamefont
  {Kienitz}, \citenamefont {Gonz{\'a}lez-F{\'e}rez},\ and\ \citenamefont
  {K{\"u}pper}}]{Trippel:PRL114:103003}%
  \BibitemOpen
  \bibfield  {author} {\bibinfo {author} {\bibfnamefont {S.}~\bibnamefont
  {Trippel}}, \bibinfo {author} {\bibfnamefont {T.}~\bibnamefont {Mullins}},
  \bibinfo {author} {\bibfnamefont {N.~L.~M.}\ \bibnamefont {M{\"u}ller}},
  \bibinfo {author} {\bibfnamefont {J.~S.}\ \bibnamefont {Kienitz}}, \bibinfo
  {author} {\bibfnamefont {R.}~\bibnamefont {Gonz{\'a}lez-F{\'e}rez}}, \ and\
  \bibinfo {author} {\bibfnamefont {J.}~\bibnamefont {K{\"u}pper}},\ }\bibfield
   {title} {\enquote {\bibinfo {title} {Two-state wave packet for strong
  field-free molecular orientation},}\ }\href {\doibase
  10.1103/PhysRevLett.114.103003} {\bibfield  {journal} {\bibinfo  {journal}
  {Phys.\ Rev.\ Lett.}\ }\textbf {\bibinfo {volume} {114}},\ \bibinfo {pages}
  {103003} (\bibinfo {year} {2015})},\ \Eprint {http://arxiv.org/abs/1409.2836}
  {arXiv:1409.2836 [physics]} \BibitemShut {NoStop}%
\bibitem [{\citenamefont {Hutzler}\ \emph {et~al.}(2012)\citenamefont
  {Hutzler}, \citenamefont {Lu},\ and\ \citenamefont
  {Doyle}}]{Hutzler:CR112:4803}%
  \BibitemOpen
  \bibfield  {author} {\bibinfo {author} {\bibfnamefont {Nicholas~R}\
  \bibnamefont {Hutzler}}, \bibinfo {author} {\bibfnamefont {Hsin-I}\
  \bibnamefont {Lu}}, \ and\ \bibinfo {author} {\bibfnamefont {John~M}\
  \bibnamefont {Doyle}},\ }\bibfield  {title} {\enquote {\bibinfo {title} {The
  buffer gas beam: An intense, cold, and slow source for atoms and
  molecules},}\ }\href {\doibase 10.1021/cr200362u} {\bibfield  {journal}
  {\bibinfo  {journal} {Chem.\ Rev.}\ }\textbf {\bibinfo {volume} {112}},\
  \bibinfo {pages} {4803--4827} (\bibinfo {year} {2012})}\BibitemShut {NoStop}%
\bibitem [{\citenamefont {Truppe}\ \emph {et~al.}(2017)\citenamefont {Truppe},
  \citenamefont {Hambach}, \citenamefont {Skoff}, \citenamefont {Bulleid},
  \citenamefont {Bumby}, \citenamefont {Hendricks}, \citenamefont {Hinds},
  \citenamefont {Sauer},\ and\ \citenamefont {Tarbutt}}]{Truppe:JModOpt65:246}%
  \BibitemOpen
  \bibfield  {author} {\bibinfo {author} {\bibfnamefont {S}~\bibnamefont
  {Truppe}}, \bibinfo {author} {\bibfnamefont {M}~\bibnamefont {Hambach}},
  \bibinfo {author} {\bibfnamefont {S~M}\ \bibnamefont {Skoff}}, \bibinfo
  {author} {\bibfnamefont {N~E}\ \bibnamefont {Bulleid}}, \bibinfo {author}
  {\bibfnamefont {J~S}\ \bibnamefont {Bumby}}, \bibinfo {author} {\bibfnamefont
  {R~J}\ \bibnamefont {Hendricks}}, \bibinfo {author} {\bibfnamefont {E~A}\
  \bibnamefont {Hinds}}, \bibinfo {author} {\bibfnamefont {B~E}\ \bibnamefont
  {Sauer}}, \ and\ \bibinfo {author} {\bibfnamefont {M~R}\ \bibnamefont
  {Tarbutt}},\ }\bibfield  {title} {\enquote {\bibinfo {title} {A buffer gas
  beam source for short, intense and slow molecular pulses},}\ }\href {\doibase
  10.1080/09500340.2017.1384516} {\bibfield  {journal} {\bibinfo  {journal}
  {J.\ Mod.\ Opt.}\ } (\bibinfo {year} {2017}),\
  10.1080/09500340.2017.1384516},\ \Eprint {http://arxiv.org/abs/1707.06291}
  {arXiv:1707.06291 [physics]} \BibitemShut {NoStop}%
\bibitem [{\citenamefont {Tobin}\ \emph {et~al.}(1987)\citenamefont {Tobin},
  \citenamefont {Sedgley}, \citenamefont {Batzer},\ and\ \citenamefont
  {Call}}]{Tobin:JVSTA5:101}%
  \BibitemOpen
  \bibfield  {author} {\bibinfo {author} {\bibfnamefont {Albert~G.}\
  \bibnamefont {Tobin}}, \bibinfo {author} {\bibfnamefont {Douglas~W.}\
  \bibnamefont {Sedgley}}, \bibinfo {author} {\bibfnamefont {Thomas~H.}\
  \bibnamefont {Batzer}}, \ and\ \bibinfo {author} {\bibfnamefont {Wayne~R.}\
  \bibnamefont {Call}},\ }\bibfield  {title} {\enquote {\bibinfo {title}
  {Evaluation of charcoal sorbents for helium cryopumping in fusion
  reactors},}\ }\href {\doibase 10.1116/1.574141} {\bibfield  {journal}
  {\bibinfo  {journal} {J. Vac. Sci. Technol. A}\ }\textbf {\bibinfo {volume}
  {5}},\ \bibinfo {pages} {101--105} (\bibinfo {year} {1987})}\BibitemShut
  {NoStop}%
\bibitem [{Note1()}]{Note1}%
  \BibitemOpen
  \bibinfo {note} {We use ``milliliter normal per minute'' as volume-
  equivalent of mass flow, with standard temperature and pressure conditions of
  \protect \ensuremath {\protect \xspace 0\protect \tmspace +\thinmuskip
  {.1667em}^\circ {}\protect \text {C}}\protect \xspace and 1.013~bar; this is
  identical to the often used literal specification ``sccm''.}\BibitemShut
  {Stop}%
\bibitem [{Com()}]{Comsol:Multiphysics:5.3}%
  \BibitemOpen
  \href@noop {} {}\bibinfo {note} {COMSOL Multiphysics v.\ 5.3.\
  \url{http://www.comsol.com}. COMSOL AB, Stockholm, Sweden}\BibitemShut
  {NoStop}%
\bibitem [{\citenamefont {Nolde}\ \emph {et~al.}(2005)\citenamefont {Nolde},
  \citenamefont {Weitzel},\ and\ \citenamefont {Western}}]{Nolde:PCCP7:1527}%
  \BibitemOpen
  \bibfield  {author} {\bibinfo {author} {\bibfnamefont {M}~\bibnamefont
  {Nolde}}, \bibinfo {author} {\bibfnamefont {KM}~\bibnamefont {Weitzel}}, \
  and\ \bibinfo {author} {\bibfnamefont {CM}~\bibnamefont {Western}},\
  }\bibfield  {title} {\enquote {\bibinfo {title} {The resonance enhanced
  multiphoton ionisation spectroscopy of ammonia isotopomers {NH$_3$},
  {NH$_2$D}, {NHD$_2$} and {ND$_3$}},}\ }\href {\doibase 10.1039/b417835c}
  {\bibfield  {journal} {\bibinfo  {journal} {Phys.\ Chem.\ Chem.\ Phys.}\
  }\textbf {\bibinfo {volume} {7}},\ \bibinfo {pages} {1527--1532} (\bibinfo
  {year} {2005})}\BibitemShut {NoStop}%
\bibitem [{\citenamefont {Patterson}\ and\ \citenamefont
  {Doyle}(2007)}]{Patterson:JCP126:154307}%
  \BibitemOpen
  \bibfield  {author} {\bibinfo {author} {\bibfnamefont {David}\ \bibnamefont
  {Patterson}}\ and\ \bibinfo {author} {\bibfnamefont {John~M.}\ \bibnamefont
  {Doyle}},\ }\bibfield  {title} {\enquote {\bibinfo {title} {Bright, guided
  molecular beam with hydrodynamic enhancement},}\ }\href {\doibase
  10.1063/1.2717178} {\bibfield  {journal} {\bibinfo  {journal} {J.\ Chem.\
  Phys.}\ }\textbf {\bibinfo {volume} {126}},\ \bibinfo {pages} {154307}
  (\bibinfo {year} {2007})}\BibitemShut {NoStop}%
\bibitem [{Note2()}]{Note2}%
  \BibitemOpen
  \bibinfo {note} {One possible loss mechanism of NH$_3$ signal is the
  formation of dimers or clusters via molecule-molecule collisions near the
  cell entrance. However, our NH$_3$ signal is observed to increase linearly
  with NH$_3$ inlet pressure and we rule out the formation of dimers or
  clusters and hence NH$_3$ loss by this mechanism.}\BibitemShut {Stop}%
\bibitem [{\citenamefont {Western}(2003-2013)}]{Western:pgopher}%
  \BibitemOpen
  \bibfield  {author} {\bibinfo {author} {\bibfnamefont {Colin~M.}\
  \bibnamefont {Western}},\ }\href@noop {} {\emph {\bibinfo {title} {{PGOPHER},
  a Program for Simulating Rotational Structure}}} (\bibinfo {year}
  {2003-2013}),\ \bibinfo {note} {{U}niversity of {B}ristol, {B}ristol, {UK},
  URL:~\url{http://pgopher.chm.bris.ac.uk}}\BibitemShut {NoStop}%
\end{thebibliography}%
\end{document}